\documentclass[12pt]{iopart}
\usepackage{graphicx}
\usepackage{iopams}
\expandafter\let\csname equation*\endcsname\relax
\expandafter\let\csname endequation*\endcsname\relax
\usepackage{amsmath,amssymb}  
\usepackage{bm}
\usepackage{mathrsfs}
\usepackage{cite} 
\usepackage[titletoc,toc,title]{appendix}

\usepackage{sidecap}
\usepackage{subfigure}

\def\bs{\bm{\sigma}}
\def\bx{\bm{\xi}}
\def\gmba{\Gamma_{b\rightarrow i}^1}
\def\gmbb{\Gamma_{b\rightarrow i}^2}
\def\gba{G_{b\rightarrow i}^1}
\def\gbb{G_{b\rightarrow i}^2}
\def\Xb{\Xi_{b\rightarrow i}}
\def\xab{\xi_i^1\xi_i^2}

\begin{document}

\title{Minimal model of permutation symmetry in unsupervised learning}
\author{Tianqi Hou}
\address{Department of Physics, the Hong Kong University of Science and Technology, Clear Water Bay, Hong Kong}
\author{K. Y. Michael Wong}
\address{Department of Physics, the Hong Kong University of Science and Technology, Clear Water Bay, Hong Kong}
\author{Haiping Huang\footnote{correspondence author}}
\address{PMI Lab, School of Physics,
Sun Yat-sen University, Guangzhou 510275, People's Republic of China}
\ead{huanghp7@mail.sysu.edu.cn}
\date{\today}

\begin{abstract}
Permutation of any two hidden units yields invariant properties in typical deep generative neural networks.
This permutation symmetry plays an important role in understanding the computation performance of a broad class of neural networks
with two or more hidden units. However, a theoretical study of the permutation symmetry is still lacking. Here, we propose 
a minimal model with only two hidden units in a restricted Boltzmann machine, which aims to address how the permutation 
symmetry affects the critical learning data size at which the concept-formation (or spontaneous symmetry breaking in physics
language) starts, and moreover semi-rigorously prove a conjecture that the critical data size is independent of the number of 
hidden units once this number is finite. Remarkably, we find that the embedded correlation between two receptive fields of hidden units
reduces the critical data size. In particular, the weakly-correlated receptive fields have the benefit of significantly reducing the minimal data size 
that triggers the transition, given less noisy data.
Inspired by the theory, we also propose an efficient fully-distributed algorithm to infer
the receptive fields of hidden units. Furthermore, our minimal model reveals that the permutation symmetry can also be spontaneously broken following the spontaneous symmetry breaking.
Overall, our results demonstrate that the unsupervised learning is a progressive combination of spontaneous symmetry breaking and
permutation symmetry breaking which are both spontaneous processes driven by data streams (observations). All these effects can be
analytically probed based on the minimal model, providing theoretical insights towards understanding unsupervised learning in a more general
context.
\end{abstract}
 \maketitle

\section{Introduction}
Unsupervised learning is defined as the process of searching for latent features in raw (unlabeled) data, and thus serves as a fundamental property of the cerebral cortex of the brain~\cite{Marr-1970,Barlow-1989}.
To understand unsupervised learning from a neural network perspective, restricted Boltzmann machine (RBM) is proposed. RBM is a two-layered neural network with one layer called
the visible layer, and the other called the hidden layer. No lateral connections exist in each layer. The connections between visible and hidden layers are called synaptic weights, representing
encoded latent features in the observed data. The process of learning the synaptic weights from the unlabeled data (also called training) mimics the unsupervised learning. RBM is thus receiving
substantial research interests both from machine learning and statistical physics communities~\cite{Hinton-2006b,Bengio-2008,Bara-2012,Huang-2015b,nips-2015,Mezard-2017,Monasson-2017,Song-2017,Decelle-2017,Sala-2017,RSB-2018}.

Training of RBM relies on the maximum likelihood principle via a gradient ascent procedure. The mean activity of each neuron or correlations between visible and hidden neurons can be estimated by either truncated Gibbs sampling~\cite{Hinton-2006b,icml-2008}
or advanced mean-field methods~\cite{Huang-2015b,nips-2015}. However, the gradient ascent method is difficult to analyze and thus not amenable for a theoretical model. Therefore, based on the
probabilistic graphical model framework, one-bit RBM where only one hidden neuron is considered was proposed to address a fundamental issue of unsupervised learning~\cite{Huang-2016b}, i.e., how many data samples are needed for a successful learning.
This work revealed a continuous spontaneous symmetry breaking (SSB) transition separating a random-guess phase from a concept-formation phase at a critical value of the amount of provided samples (data size)~\cite{Huang-2017}, which is similar to the retarded learning phase transition
observed in a generalized Hopfield model of pattern learning~\cite{Remi-2011}. This conclusion is later generalized to RBM with generic priors~\cite{Barra-2017,Barra-2018}, and synapses of ternary values~\cite{Huang-2018}.
However, it is still challenging to handle the case of multiple hidden neurons from the perspective of understanding the learning process as a phase transition. In the presence of multiple hidden neurons,
permutation symmetry appears, i.e., the model of the observed data is invariant with respect to exchange of arbitrary two hidden neurons. In addition, the permutation symmetry is a common feature in many modern neural network architectures~\cite{DL-2016}. Therefore, understanding 
how the permutation symmetry affects the concept-formation process is important, which may provide us core mechanisms of unsupervised learning.

Here, we propose a minimal model of the permutation symmetry in unsupervised learning, based on mean-field approximations. We show that it is possible to theoretically understand the permutation symmetry using
physics approximations. To be more precise, we consider a RBM with two hidden neurons, and 
embed a latent feature that generates a certain number of data samples through Gibbs samplings of the original model~\cite{Huang-2016b}. Then, the data samples are learned by a theory-inspired algorithm, and finally
the learned synaptic weights (a latent feature vector) are compared with the embedded ones, to test whether SSB applies to the minimal model, and in addition
investigate key factors affecting the critical data size for learning and moreover how the permutation symmetry affects the learning process. 
We first apply the cavity approximation in statistical mechanics of disordered systems~\cite{cavity-2001} to derive the learning algorithm from a Bayesian inference perspective, whose
computation performances in single instances of the model are then predicted by a replica theory. This theory introduces many copies of the original model, and the interaction between any two copies is characterized by a set of
self-consistent mean-field equations, from which the critical data size for learning is determined, and moreover whether the permutation symmetry can be spontaneously broken is clarified.

\begin{figure}
\centering
     \includegraphics[bb=79 598 393 775,scale=0.7]{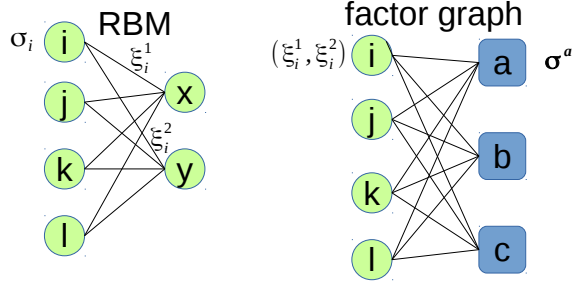}
  \caption{
  (Color online) A schematic illustration of the minimal model. $N=4$ in this example (say, $i,j,k$ and $l$). (Left panel) The original model with only two hidden neurons (say, $x$ and $y$).
  (Right panel) The corresponding factor graph where the data node is indicated by a square, and the paired-synapses (feature vector) is indicated
  by a circle. In this example, $M=3$ (say, $a,b$ and $c$). The circle is an augmented version of single synapse considered in the one-bit RBM~\cite{Huang-2016b}.
  }\label{rbm2}
\end{figure}

\section{Model definition and mean-field methods}
\subsection{Minimal model of permutation symmetry}
\label{minm}
In this study, we use the RBM defined above with two hidden neurons (Fig.~\ref{rbm2}) to learn embedded features in input data samples, which are raw unlabeled data. Each data sample is
specified by an Ising-like spin configuration $\bs=\{\sigma_i=\pm1\}_{i=1}^{N}$ where $N$ is the input dimensionality.
A collection of $M$ samples is denoted as $\{\bs^a\}_{a=1}^{M}$. Synaptic values connecting visible and hidden neurons are characterized by $\bx$, where each component takes a binary value ($\pm1$) as well. Because of two hidden neurons, 
$\bx=(\bx^1,\bx^2)$ where the superscript indicates the hidden neuron's index. $\bx^1$ and $\bx^2$ are also called receptive fields of the first and second hidden neurons, respectively. Statistical properties of
this RBM are thus described by the Boltzmann distribution~\cite{Huang-2015b}
\begin{equation}
\label{BoltM}
 P(\bs)=\frac{1}{Z(\bx)}\cosh(\beta X)\cosh(\beta Y),
\end{equation}
where $X=\frac{1}{\sqrt{N}}\bx^1\cdot\bs$, $Y=\frac{1}{\sqrt{N}}\bx^2\cdot\bs$, and $Z(\bx)$ is the partition function depending on the feature $\bx$.
Note that the two hidden neurons' activities ($\pm1$) have been marginalized out. The scaling factor $\frac{1}{\sqrt{N}}$ ensures that the argument of the hyperbolic cosine function is of the order of unity. $\bs$ can be arbitrary one of the $M$ samples.
When the embedded feature is randomly generated, the inverse-temperature $\beta$ tunes the noise level of generated data samples from the feature. Clearly, the data distribution is invariant with
respect to (w.r.t) the exchange of the hidden neurons, which is called the permutation symmetry in this paper. The required number of hidden neurons to yield this symmetry is at least two, therefore, this setup defines a minimal
model to study the permutation symmetry in unsupervised learning.

In this model, the embedded feature follows the distribution $P(\bx)=P(\bx^1)P(\bx^2|\bx^1)$ in which $P(\bx^1)=\prod^N_{i=1}\left[\frac{1}{2}\delta(\xi_i^1-1)+\frac{1}{2}\delta(\xi^1_i+1)\right]$ together with
$P(\bx^2|\bx^1)=\prod_{i=1}^N\left[p_{\rm d}\delta(\xi^2_i=-\xi_i^1)+(1-p_{\rm d})\delta(\xi_i^2=\xi_i^1)\right]$, where $p_{\rm d}$ controls the fraction of components taking different values
in the two feature maps associated with the two hidden neurons.

Given the $M$ data samples, one gets the posterior probability of the embedded feature according to the Bayes' rule:
\begin{equation}\label{Pobs}
\begin{split}
 &P(\bx|\{\bs^{a}\}_{a=1}^{M})=\frac{\prod_{a}P(\bs^{a}|\bx)}{\sum_{\bx}\prod_{a}P(\bs^{a}|\bx)}\\
 &=\frac{1}{\Omega}\prod_{a}\frac{1}{Z(\bx^1,\bx^2)}\cosh\left(\frac{\beta}{\sqrt{N}}\bx^1\cdot\bs^{a}\right)\cosh\left(\frac{\beta}{\sqrt{N}}\bx^2\cdot\bs^{a}\right),
\end{split}
 \end{equation}
where $\Omega$ is the partition function of the minimal model. For simplicity,
a uniform prior for $\bx$ is assumed, i.e., we have no prior knowledge about $\bx$, although there may exist correlations between two feature maps. In addition, we use the same temperature as that used to generate data. Because we do not 
use the true prior $\prod_iP_i(\xi^1_i,\xi^2_i|p_{{\rm d}})$, the current setting does not require the value of $p_{{\rm d}}$, thereby is not the Bayes-optimal setting which corresponds
to Nishimori condition in physics~\cite{Huang-2017}. Therefore, using the uniform prior is more computationally challenging. We leave a detailed analysis of the Bayes-optimal setting in a future work.

One obstacle to compute the posterior probability is the nested partition function $Z(\bx^1,\bx^2)$. Fortunately, this partition function can be simplified in the large-$N$
limit. More precisely,
\begin{equation}\label{Zapp}
\begin{split}
 &Z(\bx^1,\bx^2)=\sum_{\bs}\cosh\left(\frac{\beta}{\sqrt{N}}\bx^1\cdot\bs\right)\cosh\left(\frac{\beta}{\sqrt{N}}\bx^2\cdot\bs\right)\\
 &=\frac{1}{2}\sum_{\bs}\left[\cosh(X+Y)+\cosh(X-Y)\right]\\
 &=\frac{1}{2}\left[\prod_i2\cosh\left(\frac{\beta}{\sqrt{N}}(\xi^1_i+\xi_i^2)\right)+\prod_i2\cosh\left(\frac{\beta}{\sqrt{N}}(\xi^1_i-\xi_i^2)\right)\right]\\
 &\simeq2^Ne^{\beta^2}\cosh(\beta^2Q),
\end{split}
 \end{equation}
 where we have used $\ln\cosh(x)\simeq\frac{x^2}{2}$ for small $x$ to arrive at the final equality, and $Q\equiv\frac{1}{N}\sum_i\xi_i^1\xi_i^2$, which is exactly the overlap between the two feature maps. To sum up, we move all the irrelevant constants into the partition function
 $\Omega$, the posterior probability can then be rewritten as
 \begin{equation}\label{Pobs2}
 P(\bx|\{\bs^{a}\}_{a=1}^{M})=\frac{1}{\Omega}\prod_{a}\frac{1}{\cosh(\beta^2Q)}\cosh\left(\frac{\beta}{\sqrt{N}}\bx^1\cdot\bs^{a}\right)\cosh\left(\frac{\beta}{\sqrt{N}}\bx^2\cdot\bs^{a}\right),
 \end{equation}
 which forms the Boltzmann distribution of our minimal model. In this paper, we consider the case of $M=\alpha N$ where $\alpha$ specifies the data (constraint) density.

\subsection{Cavity approximation to handle the posterior probability}
\label{cavity}
In what follows, we compute the maximizer of the posterior marginals (MPM) estimator $(\hat{\xi}^1_i,\hat{\xi}_i^2) = \arg\max_{\xi^1_i,\xi_i^2}P_i(\xi_i^1,\xi_i^2)$~\cite{Huang-2017}, where the feature map of each hidden neuron is 
combined and the prediction is thus the augmented version of the inferred feature vector in the one-bit RBM~\cite{Huang-2016b}.
Hence, the task is to compute marginal probabilities, i.e., $P_i(\xi_i^1,\xi_i^2)$, which is intractable due to the interaction among data constraints (the product over $a$ in Eq.~(\ref{Pobs2})). However, by mapping
the original model (Eq.~(\ref{Pobs2})) onto a graphical model (Fig.~\ref{rbm2}), where data constraints and paired-synapses are treated respectively as factor (data) nodes and variable nodes, one can estimate the 
marginal probability by running a message passing iteration among factor and variable nodes, as we shall explain below.
The key assumption is that the paired-synapses on the graphical model are weakly correlated, which is called the Bethe approximation~\cite{MM-2009} in physics.

We first define a cavity
probability $P_{i\rightarrow a}(\xi_i^1,\xi_i^2)$ with the data node $a$ removed. Under the weak correlation assumption, $P_{i\rightarrow a}(\xi_i^1,\xi_i^2)$
obeys a self-consistent equation:
\begin{subequations}\label{bp0}
\begin{align}
P_{i\rightarrow a}(\xi_{i}^1,\xi_i^2)&=\frac{1}{Z_{i\rightarrow a}}
\prod_{b\in\partial i\backslash
a}\mu_{b\rightarrow i}(\xi_{i}^1,\xi_i^2),\label{bp2}\\
\begin{split}
\mu_{b\rightarrow i}(\xi_{i}^1,\xi_i^2)&=\sum_{\bx\backslash\xi_i^1,\xi_i^2}\frac{1}{\cosh\left(\beta^2Q_c+\frac{\beta^2}{N}\xi_i^1\xi_i^2\right)}\cosh\left(\beta X_b+\frac{\beta}{\sqrt{N}}\xi_i^1\sigma_i^b\right)\cosh\left(\beta Y_b+\frac{\beta}{\sqrt{N}}\xi_i^2\sigma_i^b\right)\\
&\times\prod_{j\in\partial
  b\backslash i}P_{j\rightarrow b}(\xi_{j}^1,\xi_j^2),\label{bp1}
\end{split}
\end{align}
\end{subequations}
where $Z_{i\rightarrow a}$ is a normalization constant,
$\partial i\backslash a$ denotes neighbors of the feature node $i$ except the
data node $a$, $\partial b\backslash i$ denotes neighbors of the
data node $b$ except the feature node $i$, and the auxiliary quantity
$\mu_{b\rightarrow i}(\xi_i^1,\xi_i^2)$ denotes the contribution from
data node $b$ to feature node $i$ given the value of
$(\xi_i^1,\xi_i^2)$~\cite{Huang-2015b,Huang-2016b}. Products in Eq.~(\ref{bp0}) result from the weak correlation assumption.
In addition, $X_b\equiv\frac{1}{\sqrt{N}}\sum_{j\neq i}\xi^1_j\sigma_j^b$, $Y_b\equiv\frac{1}{\sqrt{N}}\sum_{j\neq i}\xi^2_j\sigma_j^b$, and the cavity version of
$Q$ is defined as $Q_c\equiv\frac{1}{N}\sum_{j\neq i}\xi_j^1\xi_j^2$, which can be further replaced by its typical value obtained by the average over the cavity probability (to be shown below). Although this is a crude approximation,
it works quite well in practice.

Still, the above self-consistent equation is intractable due to the summation to estimate $\mu_{b\rightarrow i}$.
Nevertheless, a careful inspection reveals that $X_b$ and $Y_b$ are approximately correlated Gaussian random variables due to the central limit theorem.
As a result, the intractable summation can be replaced by an integral which is easy to calculate in this model. We just need to compute the following mean, variance and covariance between these random variables.
\begin{subequations}\label{cov}
\begin{align}
G_{b\rightarrow i}^1&=\frac{1}{\sqrt{N}}\sum_{j\neq i}\sigma_j^bm_{j\rightarrow b}^1,\\
G_{b\rightarrow i}^2&=\frac{1}{\sqrt{N}}\sum_{j\neq i}\sigma_j^bm_{j\rightarrow b}^2,\\
\Gamma_{b\rightarrow i}^1&=\frac{1}{N}\sum_{j\neq i}\left(1-(m_{j\rightarrow b}^1)^2\right),\\
\Gamma_{b\rightarrow i}^2&=\frac{1}{N}\sum_{j\neq i}\left(1-(m_{j\rightarrow b}^2)^2\right),\\
\Xi_{b\rightarrow i}&=\frac{1}{N}\sum_{j\neq i}\left(q_{j\rightarrow b}-m_{j\rightarrow b}^1m_{j\rightarrow b}^2\right),
\end{align}
\end{subequations}
where $G$ and $\Gamma$ denotes the mean and variance of the Gaussian random variable respectively, and the last quantity denotes the covariance between $X_b$ and $Y_b$. 
The cavity magnetization is defined as $m_{j\rightarrow b}^{1,2}=\sum_{\xi_j^1,\xi_j^2}\xi_j^{1,2}P_{j\rightarrow b}(\xi_j^1,\xi_j^2)$, and
the cavity correlation is defined as $q_{j\rightarrow b}=\sum_{\xi_j^1,\xi_j^2}\xi_j^1\xi_j^2P_{j\rightarrow b}(\xi_j^1,\xi_j^2)$. Finally, using the above parameters of the correlated Gaussian distribution, we rewrite $\mu_{b\rightarrow i}(\xi_i^1,\xi_i^2)$
as
\begin{equation}\label{mu}
\begin{split}
 \mu_{b\rightarrow i}(\xi_i^1,\xi_i^2)&=\frac{1}{\cosh\left(\beta^2Q_{b\rightarrow i}
 +\frac{\beta^2}{N}\xi^1_i\xi^2_i\right)}\iint DxDy\cosh\left(\beta\sqrt{\gmba}x+\beta\gba+\frac{\beta}{\sqrt{N}}\xi^1_i\sigma_i^b\right)\\
 &\times\cosh\left(\beta\sqrt{\gmbb}(\psi x+\sqrt{1-\psi^2}y)+\beta\gbb+\frac{\beta}{\sqrt{N}}\xi^2_i\sigma_i^b\right),
\end{split}
 \end{equation}
where $Dx\equiv\frac{e^{-x^2/2}dx}{\sqrt{2\pi}}$, $\psi=\frac{\Xb}{\sqrt{\gmba\gmbb}}$, and $Q_{b\rightarrow i}=\frac{1}{N}\sum_{j\neq i}q_{j\rightarrow b}$ stemming from $Q_c$ in Eq.~(\ref{bp1}) replaced by its
cavity mean. The above integral representation of 
$\mu_{b\rightarrow i}(\xi^1_i,\xi^2_i)$ can be analytically estimated; for convenience, we define $u_{b\rightarrow i}(\xi_i^1,\xi_i^2)\equiv\ln\mu_{b\rightarrow i}(\xi_i^1,\xi^2_i)$.
It is easy to show that
\begin{equation}\label{ub}
\begin{split}
 u_{b\rightarrow i}(\xi_i^1,\xi_i^2)&=\frac{\beta^2\gmbb(1-\psi^2)}{2}-\ln\left(2\cosh\Bigl(\beta^2Q_{b\rightarrow i}+\frac{\beta^2\xab}{N}\Bigr)\right)+\frac{\beta^2}{2}\left(\sqrt{\gmba}+\sqrt{\gmbb}\psi\right)^2\\
 &+\ln\cosh\left(\beta\gba+\beta\gbb+\frac{\beta}{\sqrt{N}}\sigma_i^b(\xi^1_i+\xi^2_i)\right)\\
 &+
 \ln\left[1+e^{-2\beta^2\sqrt{\gmba\gmbb}\psi}\frac{\cosh\left(\beta\gba-\beta\gbb+\frac{\beta}{\sqrt{N}}\sigma_i^b(\xi^1_i-\xi^2_i)\right)}{\cosh\left(\beta\gba+\beta\gbb+\frac{\beta}{\sqrt{N}}\sigma_i^b(\xi^1_i+\xi^2_i)\right)}\right].
\end{split}
 \end{equation}
 
To close the iteration equation, we need to compute the cavity magnetization and correlation as follows:
\begin{subequations}\label{bp3}
\begin{align}
m_{i\rightarrow a}^1&=\frac{\sum_{\xi^1=\pm1,\xi^2=\pm1}\xi^1e^{\sum_{b\in\partial i\backslash a}u_{b\rightarrow i}(\xi^1,\xi^2)}}{\sum_{\xi^1=\pm1,\xi^2=\pm1}e^{\sum_{b\in\partial i\backslash a}u_{b\rightarrow i}(\xi^1,\xi^2)}},\label{bp3a}\\
m_{i\rightarrow a}^2&=\frac{\sum_{\xi^1=\pm1,\xi^2=\pm1}\xi^2e^{\sum_{b\in\partial i\backslash a}u_{b\rightarrow i}(\xi^1,\xi^2)}}{\sum_{\xi^1=\pm1,\xi^2=\pm1}e^{\sum_{b\in\partial i\backslash a}u_{b\rightarrow i}(\xi^1,\xi^2)}},\label{bp3b}\\
q_{i\rightarrow a}&=\frac{\sum_{\xi^1=\pm1,\xi^2=\pm1}\xi^1\xi^2e^{\sum_{b\in\partial i\backslash a}u_{b\rightarrow i}(\xi^1,\xi^2)}}{\sum_{\xi^1=\pm1,\xi^2=\pm1}e^{\sum_{b\in\partial i\backslash a}u_{b\rightarrow i}(\xi^1,\xi^2)}}.\label{bp3c}
\end{align}
\end{subequations}
$m_{i\rightarrow a}^{1,2}$ can be interpreted as the message passing from feature node $i$ to data node $a$ ($q_{i\rightarrow a}$ is also similarly interpreted), while $u_{b\rightarrow i}$ can be
interpreted as the message passing from data node $b$ to feature node $i$. 

If the weak correlation assumption is
self-consistent, starting from randomly initialized messages, the learning equations will converge
to a fixed point corresponding to a thermodynamically dominant minimum of the Bethe free energy function~\cite{Yedidia-2005}, which is given by $-\beta f_{{\rm Bethe}}=\frac{1}{N}\sum_{i}\Delta f_i-\frac{N-1}{N}\sum_{a}\Delta f_a$. 
The free energy contributions of variable node and data node are given respectively by:
\begin{subequations}\label{fifa}
\begin{align}
\Delta f_i&=\ln\sum_{\xi^1_i,\xi^2_i}\prod_{b\in\partial i}\mu_{b\rightarrow i}(\xi^1_i,\xi^2_i),\\
\begin{split}
\Delta f_a&=\frac{\beta^2\Gamma_a^2(1-\tilde{\psi}^2)}{2}-\ln\left(2\cosh(\beta^2Q_{a})\right)+\frac{\beta^2}{2}\left(\sqrt{\Gamma_a^1}+\sqrt{\Gamma_a^2}\tilde{\psi}\right)^2\\
 &+\ln\cosh\left(\beta G_{a}^1+\beta G_a^2\right)
 +
 \ln\left[1+e^{-2\beta^2\Xi_a}\frac{\cosh\left(\beta G_a^1-\beta G_a^2\right)}{\cosh\left(\beta G_a^1+\beta G_a^2\right)}\right],
 \end{split}
\end{align}
\end{subequations}
where $\tilde{\psi}=\frac{\Xi_a}{\sqrt{\Gamma_a^1\Gamma_a^2}}$. The forms of $\Gamma_a^{1,2}$, $G_a^{1,2}$, $Q_a$ and $\Xi_a$ are similar to their cavity counterparts (e.g., in Eq.~(\ref{ub})), but with the only difference that the node $i$'s contribution is not excluded.
Once the iteration converges, the MPM estimator predicts that $\hat{\xi}^1_i={\rm sgn}(m_{i}^1)$ and $\hat{\xi}^2_i={\rm sgn}(m_i^2)$, where the full (non-cavity) magnetization $m_i^{1,2}$ is computed taking into account all contributions of adjacent data nodes to the node $i$ (see Eq.~(\ref{bp3}),
and the symbol $\backslash a$ is thus removed).

\subsection{Replica theory of the minimal model}
\label{theor}
 
 To have an analytic argument about the critical threshold for spontaneous symmetry breaking, we calculate the free energy
 in the thermodynamic limit using the replica method. Instead of calculating a disorder average of $\ln\Omega$, the replica method computes the disorder average of
an integer power of $\Omega$, i.e., $\left<\Omega^n\right>$. In physics, this corresponds to preparing $n$ replicas of the original system;
then the rescaled free energy density (multiplied by $-\beta$) can be obtained as~\cite{Huang-2017}
\begin{equation}\label{replica}
 -\beta f=\lim_{n\rightarrow 0,N\rightarrow\infty}\frac{\ln\left<\Omega^n\right>}{nN},
\end{equation}
where the limits of $N\rightarrow\infty$ and $n\rightarrow0$ have been exchanged, such that the thermodynamic limit can be taken first
for applying the Laplace's method or saddle-point analysis~\cite{Nishimori-2001}, and the disorder average is taken over all possible samplings (data) and the random realizations of the true feature vector.
The explicit form of $\left<\Omega^n\right>$ reads
\begin{equation}\label{Zn}
\begin{split}
 \left<\Omega^n\right>&=\sum_{\{\bs^a,\bx^{{\rm true}}\}}\prod_i\left[P(\xi_i^{1,{\rm true}},\xi_i^{2,{\rm true}})\right]\prod_a\frac{\cosh\left(\frac{\beta}{\sqrt{N}}\bx^{1,{\rm true}}\cdot\bs^a\right)\cosh\left(\frac{\beta}{\sqrt{N}}\bx^{2,{\rm true}}\cdot\bs^a\right)}
 {2^Ne^{\beta^2}\cosh(\beta^2q)}\\
 &\times\sum_{\{\bx^{1,\gamma},\bx^{2,\gamma}\}}\prod_{a,\gamma}\frac{\cosh\left(\frac{\beta}{\sqrt{N}}\bx^{1,\gamma}\cdot\bs^a\right)\cosh\left(\frac{\beta}{\sqrt{N}}\bx^{2,\gamma}\cdot\bs^a\right)}{\cosh(\beta^2R^\gamma)},
\end{split}
 \end{equation}
where $\gamma$ indicates the replica index, $\bx^{{\rm true}}\equiv(\bx^{1,{\rm true}},\bx^{2,{\rm true}})$, $q\equiv\frac{1}{N}\bx^{1,{\rm true}}\cdot\bx^{2,{\rm true}}$, and $R^\gamma\equiv\frac{1}{N}\bx^{1,\gamma}\cdot\bx^{2,\gamma}$.
Note that $q$ is pre-determined and used to generate the random true feature maps, as also defined in section~\ref{minm}.
We leave the technical details to~\ref{app1}, and give the final result here. The free energy function reads,
\begin{equation}\label{freeRBM2}
 \begin{split}
 -\beta f_{{\rm RS}}&=-R\hat{R}-T_1\hat{T}_1-\tau_1\hat{\tau}_1-T_2\hat{T}_2-\tau_2\hat{\tau}_2+\frac{\hat{q}_1(q_1-1)}{2}+\frac{\hat{q}_2(q_2-1)}{2}+\frac{r\hat{r}}{2}\\
 &+\alpha\beta^2\left(1-\frac{q_1+q_2}{2}\right)-\alpha\ln\Bigl(2\cosh(\beta^2R)\Bigr)+\int D\mathbf{z}\left[\ln Z_{{\rm eff}}\right]_{\xi^{1,{\rm true}},\xi^{2,{\rm true}}}\\
 &+\frac{\alpha e^{-\beta^2}}{\cosh(\beta^2q)}
 \int D\mathbf{t}\cosh(\beta t_0)\cosh(\beta qt_0+\beta\sqrt{1-q^2}x_0)\ln Z_{{\rm E}},
 \end{split}
\end{equation}
where $[\cdot]_{\xi^{1,{\rm true}},\xi^{2,{\rm true}}}$ means an average w.r.t $P(\xi^{1,{\rm true}},\xi^{2,{\rm true}})$, $D\mathbf{z}\equiv Dz_1Dz_2Dz_3$ (a standard Gaussian measure vector, as defined below Eq.~(\ref{mu})), and similarly $D\mathbf{t}\equiv Dt_0Dx_0DuDu'$. ${\rm RS}$ means the replica symmetry assumption we used to get 
the final result. This assumption implies that the order parameter (various kinds of overlaps, explicitly defined below) does not rely on its specific replica index.
We assume that this assumption is able to describe the system as we shall show it leads to consistent predictions verified in algorithmic results of single instances.
The auxiliary quantities $Z_{{\rm eff}}$ and $Z_{{\rm E}}$ are defined as follows,
\begin{subequations}\label{rbmReplica1}
\begin{align}
Z_{{\rm eff}}&=\sum_{\xi^1,\xi^2}e^{b_1\xi^1+b_2\xi^2+b_3\xi^1\xi^2},\\
b_1&=\hat{T}_1\xi^{1,{\rm true}}+\hat{\tau}_2\xi^{2,{\rm true}}+\sqrt{\hat{q}_1-\hat{r}/2}z_1+\sqrt{\hat{r}/2}z_3,\\
b_2&=\hat{T}_2\xi^{2,{\rm true}}+\hat{\tau}_1\xi^{1,{\rm true}}+\sqrt{\hat{q}_2-\hat{r}/2}z_2+\sqrt{\hat{r}/2}z_3,\\
b_3&=\hat{R}-\hat{r}/2,\\
Z_{{\rm E}}&=e^{\beta^2(R-r)}\cosh(\beta\Lambda_{+})+e^{-\beta^2(R-r)}\cosh(\beta\Lambda_{-}),\\
\Lambda_{+}&=(T_1+\tau_1)t_0+\frac{1}{\sqrt{1-q^2}}(T_2+\tau_2-q(T_1+\tau_1))x_0+(B+\frac{r-A}{B})u+Ku',\\
\Lambda_{-}&=(T_1-\tau_1)t_0+\frac{1}{\sqrt{1-q^2}}(\tau_2-T_2-q(T_1-\tau_1))x_0+(B-\frac{r-A}{B})u-Ku',
\end{align}
\end{subequations}
where $A\equiv T_1\tau_1+\frac{(\tau_2-T_1q)(T_2-\tau_1q)}{1-q^2}$, $B\equiv\sqrt{q_1-T_1^2-\frac{(\tau_2-T_1q)^2}{1-q^2}}$, and
$K\equiv\sqrt{q_2-\tau_1^2-\frac{(T_2-\tau_1q)^2}{1-q^2}-\frac{(r-A)^2}{B^2}}$.

The associated (non-conjugated) saddle-point equations are expressed as
\begin{subequations}\label{rbmReplica2}
\begin{align}
T_1&=\left[\xi^{1,{\rm true}}\left<\xi^1\right>\right]_{\mathbf{z},\xi^{1,{\rm true}},\xi^{2,{\rm true}}},\\
T_2&=\left[\xi^{2,{\rm true}}\left<\xi^2\right>\right]_{\mathbf{z},\xi^{1,{\rm true}},\xi^{2,{\rm true}}},\\
q_1&=\left[\left<\xi^1\right>^2\right]_{\mathbf{z},\xi^{1,{\rm true}},\xi^{2,{\rm true}}},\\
q_2&=\left[\left<\xi^2\right>^2\right]_{\mathbf{z},\xi^{1,{\rm true}},\xi^{2,{\rm true}}},\\
\tau_1&=\left[\xi^{1,{\rm true}}\left<\xi^2\right>\right]_{\mathbf{z},\xi^{1,{\rm true}},\xi^{2,{\rm true}}},\\
\tau_2&=\left[\xi^{2,{\rm true}}\left<\xi^1\right>\right]_{\mathbf{z},\xi^{1,{\rm true}},\xi^{2,{\rm true}}},\\
R&=\left[\left<\xi^1\xi^2\right>\right]_{\mathbf{z},\xi^{1,{\rm true}},\xi^{2,{\rm true}}},\\
r&=\left[\left<\xi^1\right>\left<\xi^2\right>\right]_{\mathbf{z},\xi^{1,{\rm true}},\xi^{2,{\rm true}}}.
\end{align}
\end{subequations}
 Note that the average w.r.t the true features can be written explicitly by definition as $P(\xi^{1,{\rm true}},\xi^{2,{\rm true}})=\frac{p_{{\rm d}}}{2}$ for
 both true components taking different values, and otherwise $P(\xi^{1,{\rm true}},\xi^{2,{\rm true}})=\frac{1-p_{{\rm d}}}{2}$. $p_{{\rm d}}$ is related to $q$ by
 $p_{{\rm d}}=\frac{1-q}{2}$. The outer average also includes the disorder average over $\mathbf{z}$. The inner average $\left<\bullet\right>$ indicates the thermal average under the partition function $Z_{{\rm eff}}$ (corresponding to a two-spin interaction Hamiltonian).
 This average is analytically tractable, e.g.,
 $\left<\xi^1\right>=\frac{1}{Z_{{\rm eff}}}\frac{\partial Z_{{\rm eff}}}{\partial b_1}=\frac{\tanh b_1+\tanh b_2\tanh b_3}{1+\tanh b_1\tanh b_2\tanh b_3}$. $\left<\xi^2\right>$ and $\left<\xi^1\xi^2\right>$ can also be similarly computed.
 
 We further comment that $T_1$ characterizes the typical overlap between inferred value of the first feature and its true counterpart, and likewise
 $T_2$ characterizes the typical overlap between the second feature and its ground truth; $q_1$ and $q_2$ characterize the sizes of the first and second feature spaces respectively;
 $R$ characterizes the correlation of the two features within the same replica, while $r$ is the correlation for different replicas; $\tau_1$ characterizes
 the typical correlation between the first true feature and the inferred value of the second feature in an arbitrary replica, and likewise $\tau_2$ characterizes
 the typical correlation between the second true feature and the inferred value of the first feature in an arbitrary replica. $\tau_1$ and $\tau_2$ are thus responsible for the permutation symmetry effect.
 Taken all together,
 $(T_1,T_2,q_1,q_2,R,r,\tau_1,\tau_2)$ forms the order parameter set of
 our model. Their exact mathematical definitions are given in the~\ref{app1}.
 
 Finally, the conjugated order parameters can also be derived from a saddle point analysis of the free energy function, and they obey the following equations:
 \begin{subequations}\label{rbmReplica3}
\begin{align}
\hat{T}_1&=\alpha\beta^2\langle\langle G_s^+\rangle\rangle,\\
\hat{T}_2&=\alpha\beta^2\langle\langle\langle G_s^-\rangle\rangle\rangle,\\
\hat{q}_1&=\alpha\beta^2\left<(G_s^+)^2\right>,\\
\hat{q}_2&=\alpha\beta^2\left<(G_s^-)^2\right>,\\
\hat{\tau}_1&=\alpha\beta^2\langle\langle G_s^-\rangle\rangle,\\
\hat{\tau}_2&=\alpha\beta^2\langle\langle\langle G_s^+\rangle\rangle\rangle,\\
\hat{R}&=\alpha\beta^2\left<G_c^-\right>-\alpha\beta^2\tanh(\beta^2R),\\
\hat{r}&=2\alpha\beta^2\left<G_s^+G_s^-\right>,
\end{align}
\end{subequations}
where the average $\left<\bullet\right>\equiv\frac{e^{-\beta^2}}{\cosh\beta^2q}\int D\mathbf{t}\cosh(\beta t_0)\cosh(\beta qt_0+\beta\sqrt{1-q^2}x_0)\bullet$,
$\langle\langle\bullet\rangle\rangle\equiv\frac{e^{-\beta^2}}{\cosh\beta^2q}\int D\mathbf{t}\sinh(\beta t_0)\cosh(\beta qt_0+\beta\sqrt{1-q^2}x_0)\bullet$, 
and $\langle\langle\langle\bullet\rangle\rangle\rangle\equiv\frac{e^{-\beta^2}}{\cosh\beta^2q}\int D\mathbf{t}\cosh(\beta t_0)\sinh(\beta qt_0+\beta\sqrt{1-q^2}x_0)\bullet$. The auxiliary quantities 
are defined as follows,
\begin{subequations}\label{rbmReplica4}
\begin{align}
G_c^-&=\frac{e^{\beta^2(R-r)}\cosh\beta\Lambda_+-e^{-\beta^2(R-r)}\cosh\beta\Lambda_{-}}{e^{\beta^2(R-r)}\cosh\beta\Lambda_++e^{-\beta^2(R-r)}\cosh\beta\Lambda_{-}},\\
G_s^+&=\frac{e^{\beta^2(R-r)}\sinh\beta\Lambda_++e^{-\beta^2(R-r)}\sinh\beta\Lambda_{-}}{e^{\beta^2(R-r)}\cosh\beta\Lambda_++e^{-\beta^2(R-r)}\cosh\beta\Lambda_{-}},\\
G_s^-&=\frac{e^{\beta^2(R-r)}\sinh\beta\Lambda_+-e^{-\beta^2(R-r)}\sinh\beta\Lambda_{-}}{e^{\beta^2(R-r)}\cosh\beta\Lambda_++e^{-\beta^2(R-r)}\cosh\beta\Lambda_{-}}.
\end{align}
\end{subequations}

To sum up, Eqs.~(\ref{rbmReplica2}) and~(\ref{rbmReplica3}) construct a closed iterative equation (detailed derivations are given in the~\ref{app2}), whose fixed point 
gives an approximate evaluation of the free energy. When all order parameters vanish (a trivial disordered state), the free energy has 
an analytic value expressed as $\alpha\beta^2+\ln4$ in agreement with $-\beta f_{\rm Bethe}$ in the same trivial state. In addition,
these saddle point equations in the case of $q=0$ can be simplified to the result of Ref.~\cite{Huang-2017} for 
unsupervised feature learning in a one-bit RBM (see the~\ref{app4}). More precisely, when the true feature maps are orthogonal, we have $\Omega=\Omega_{{\rm one-bit-RBM}}^2$, thus the free
energy is two times as large as that of one-bit RBM.

Meanwhile,
the converged order parameters from Eqs.~(\ref{rbmReplica2}) and~(\ref{rbmReplica3}), especially $T_1$ and $T_2$ can be compared with the algorithmic results, and can also be used to analytically derive the critical threshold 
$\alpha_c$ for unsupervised learning in this permutation-symmetry model. When the data size is not sufficient, we expect that the order parameters vanish, and in the small order-parameter limit,
$T_1\simeq\hat{T}_1+q\hat{\tau}_2$, $\tau_2\simeq\hat{\tau}_2+\hat{T}_1q$, $\hat{T}_1\simeq\alpha\beta^4[T_1+\tau_2\tanh(\beta^2q)]$, and $\hat{\tau}_2\simeq\alpha\beta^4[T_1\tanh(\beta^2q)+\tau_2]$.
Based on these four equations, it is easy to show that the critical learning threshold is given by
\begin{equation}\label{thre}
 \alpha_c=\frac{\beta^{-4}}{1+q\tanh(\beta^2q)+|\tanh(\beta^2q)+q|}.
\end{equation}

\begin{figure}
\centering
   \includegraphics[bb=49 206 537 582,scale=0.45]{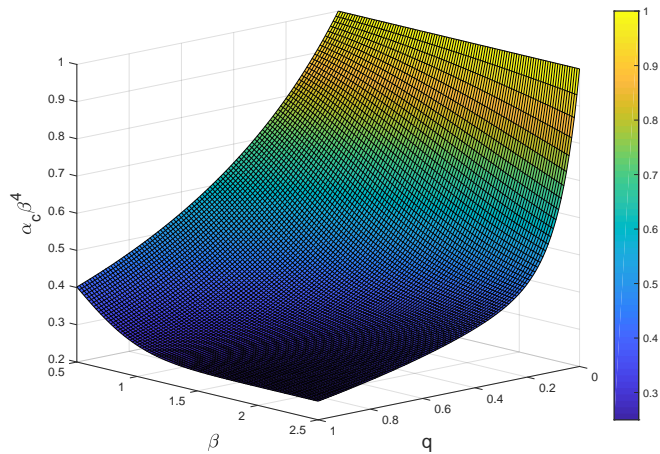}
  \caption{
  (Color online) The critical value of data size (Eq.~(\ref{thre})) for learning in our minimal model, as a function of the noise level $\beta$ as well
  as the correlation level $q$.
  }\label{figthre}
\end{figure}

As expected, the critical value for our current model does not depend on
the sign of $q$. A detailed derivation of $\alpha_c$ is given in the~\ref{app3}. In the correlation-free limit $q=0$,
the known result of the one-bit RBM, $\alpha_c=\beta^{-4}$~\cite{Huang-2016b,Huang-2017} is recovered. Thus, we first theoretically prove the conjecture made by empirical observations in Ref.~\cite{Barra-2017} that once the
number of hidden neurons is finite (not proportional to $N$), the critical learning threshold does not change with this finite number! However, Ref.~\cite{Barra-2017} overlooks the effect of 
the potential correlation across true hidden features, which indeed affects the learning threshold. Remarkably, this effect, more natural than the ideal correlation-free case, is clearly captured by our theory (Eq.~(\ref{thre})).
We show this effect in Fig.~\ref{figthre}. We can see that for a fixed noise level, increasing $q$ has the effect of decreasing the critical data size. That is, if the data set is created by
strongly correlated receptive fields (feature maps), then the learning is relatively easy, or fewer samples are sufficient to trigger the phase transition of concept-formation.
Conversely, if $q$ is zero, the data is generated from independent feature maps, then a successful learning requires a much more larger
dataset. We also observe that for a large $\beta$ (less noisy data), a small correlation level $q$ can already significantly
reduce the minimal data size that triggers learning, as verified by the fact that around the small $q$ region the larger the $\beta$ is, the sharper the surface becomes. 

\begin{figure}
\centering
     \includegraphics[bb=0 0 444 334,scale=0.57]{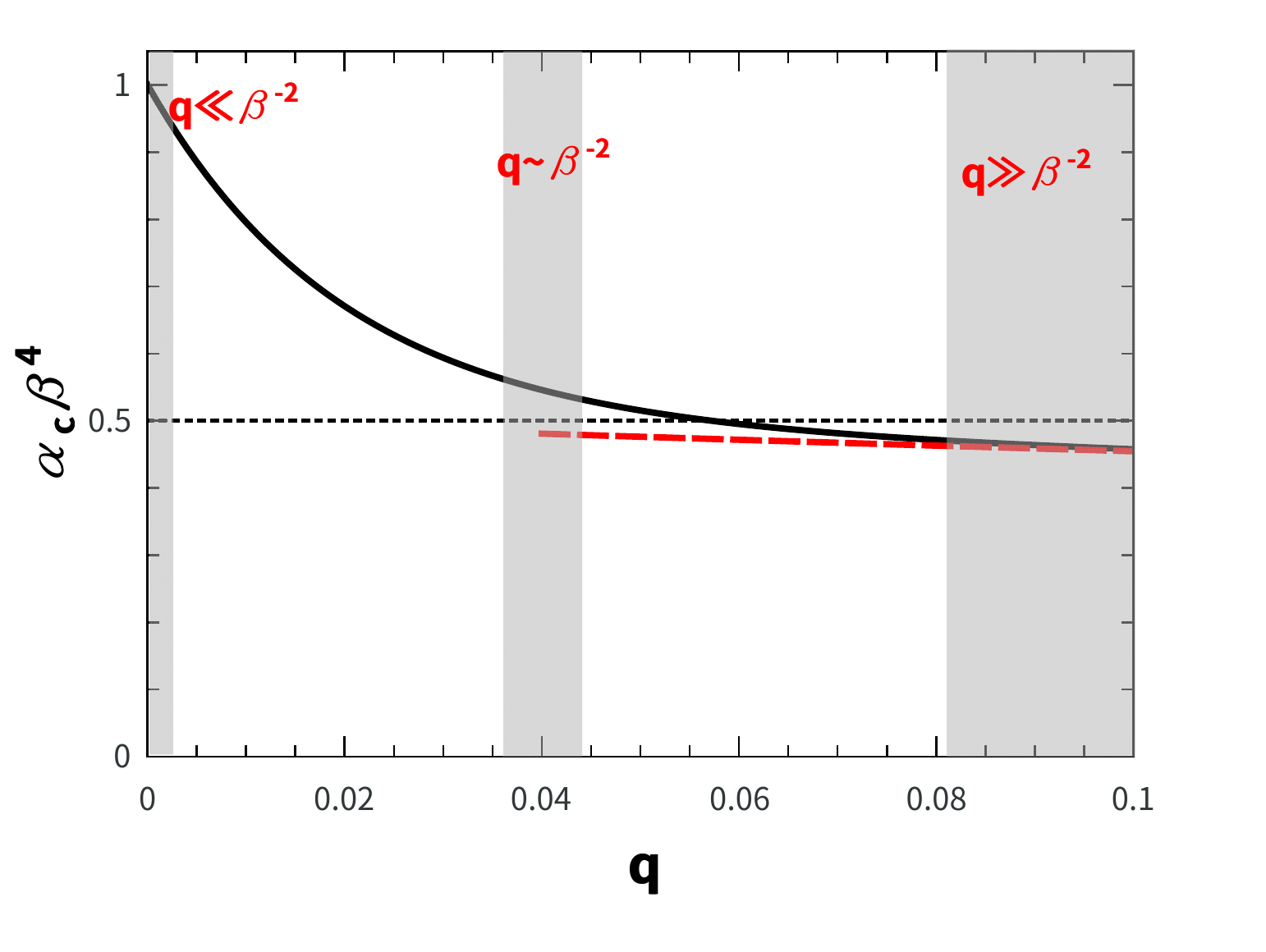}
  \caption{
  (Color online) The critical value of data size (Eq.~(\ref{thre}))
  as a function of the noise level $q$. We consider the weak-feature-correlation limit at different orders of magnitude compared with $\beta^{-2}$. We use the value of $\beta=5$ for an example. 
  The dashed line indicates the third case of Eq.~(\ref{alimit}). 
  }\label{thre0}
\end{figure}

Next, we analyze two interesting limits implied by the critical threshold equation (Eq.~(\ref{thre})).
In the limit $|q|\rightarrow1$, $\alpha_c\rightarrow\frac{1}{4}\beta^{-4}$ provided that $\beta$ is relatively large such that $\tanh\beta^2\simeq1$.
The second case is another limit $|q|\rightarrow0$, i.e., $q$ takes a small value but not zero, implying that a weak correlation among feature maps is maintained. 
Depending on the order of magnitude of $q$, we have the following result given a relatively large $\beta$:
\begin{equation}\label{alimit}
 \lim_{\beta\rightarrow\infty}\alpha_c\beta^4  = \begin{cases}
 1& \textrm{if}\, |q|\ll\beta^{-2},\\
\frac{1}{1+|\tanh q_0|} & \textrm{if}\, q=q_0\beta^{-2}\,\textrm{or $|q|\sim\beta^{-2}$},\\
\frac{1}{2(1+|q|)} & \textrm{if}\, |q|\gg\beta^{-2}.
\end{cases}
\end{equation}
Note that $\infty$ means any large value of $\beta$ such that $\tanh\beta\simeq1$ rather than a definite value of infinity. Eq.~(\ref{alimit}) reveals that
once the two feature maps are weakly correlated, the minimal learning data size for a transition can be further (or even significantly) reduced compared to the correlation-free case, 
particularly in the case that
$q$ is not very small but still larger than the order of magnitude set by $\beta^{-2}$. We show this result in Fig.~\ref{thre0}. 

We thereby have a \textit{significant hypothesis} for the triggering of concept-formation that a bit large (compared with $\beta^{-2}$) yet still small value of 
the correlation level is highly favored for unsupervised learning from
a dataset of smaller size (compared with the correlation-free case). 
Regularization techniques such as locally enforcing feature orthogonality~\cite{ICLR-17} and dropping some connections during training~\cite{dropcon} have been introduced to deep learning. Weakly-correlated receptive fields are also favored from the 
perspective of neural computation, since the redundancy among synaptic weights is reduced and thus different feature detectors inside the network can encode efficiently stimuli features rather than capturing noise in the data.
A similar decorrelation in hidden activities was recently theoretically analyzed in feedforward neural networks~\cite{Huang-2018b}.
We hope our theoretical prediction can be verified in specific machine learning tasks, and even in neuroscience experiments where
the relationship amongst the minimal data size for learning, the correlation level of synapses (or receptive fields) and the noise level in stimuli can be jointly established.
Therefore, from the Bayesian learning perspective,
the correlated-feature-map case yields a much lower threshold of phase transition towards the concept formation, in comparison with the correlation-free case~
\cite{Huang-2016b,Huang-2017,Barra-2017}.

Overall, the prediction quantitatively captures the learning behavior in both uncorrelated and correlated settings.
In next section, we shall further verify this conclusion by extensive numerical simulations on single instances of the minimal model.

\section{Results and discussion}
\label{disc}
In this section, we study how the permutation symmetry between two hidden neurons affects the learning process, i.e., the spontaneous symmetry breaking transition of the concept-formation during unsupervised
learning. We focus on whether the replica theory predicts the learning threshold related to the continuous transition. The learning threshold can be estimated from
the message passing algorithmic results on single instances of the minimal model. We first randomly generate true feature maps with a pre-determined correlation level specified by $q$.
The true feature maps are then used to generate $M$ Monte-Carlo samples through a Gibbs sampling procedure~\cite{Huang-2016b} according to Eq.~(\ref{BoltM}).
Finally, these samples are used as a quenched disorder for the Bayesian inference of the true feature maps~\cite{Huang-2017}. An overlap with the ground truth is computed and compared with 
the replica prediction. In fact, for comparison, we define the overlap~\cite{Huang-2017}, e.g., $T_1^{\rm{MP}}=\frac{1}{N}\sum_i\xi_i^{\rm{true}}m_i^{1}$ where $\rm{MP}$ means message passing, and $m_i^1$ takes into
account the thermal average (uncertainty about the ground truth). Other overlaps can be similarly defined.

\begin{figure}
\centering
\subfigure[]{
     \includegraphics[bb=1 11 364 311,width=.42\textwidth]{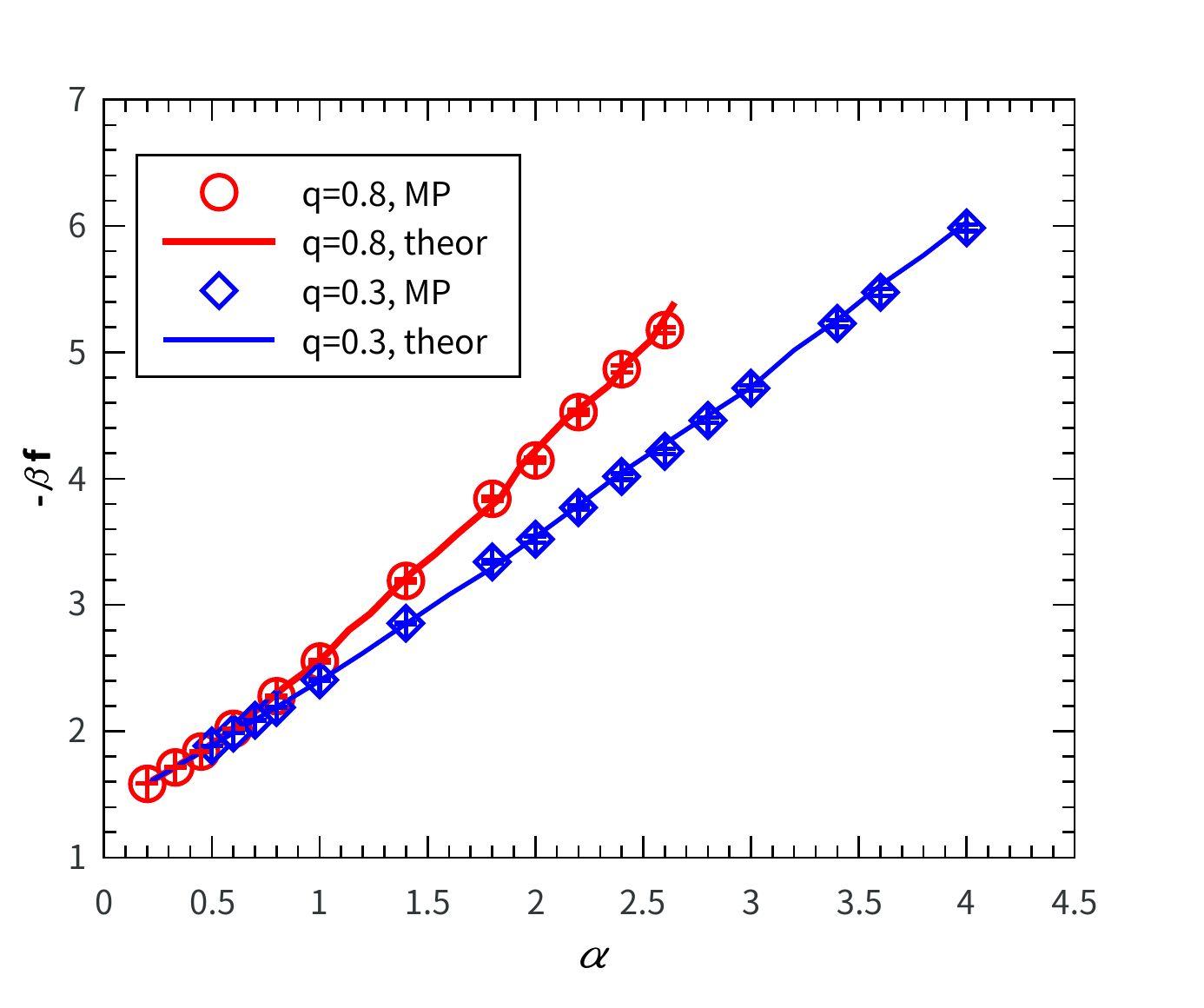}}
     \subfigure[]{\includegraphics[bb=0 0 400 307,width=.45\textwidth]{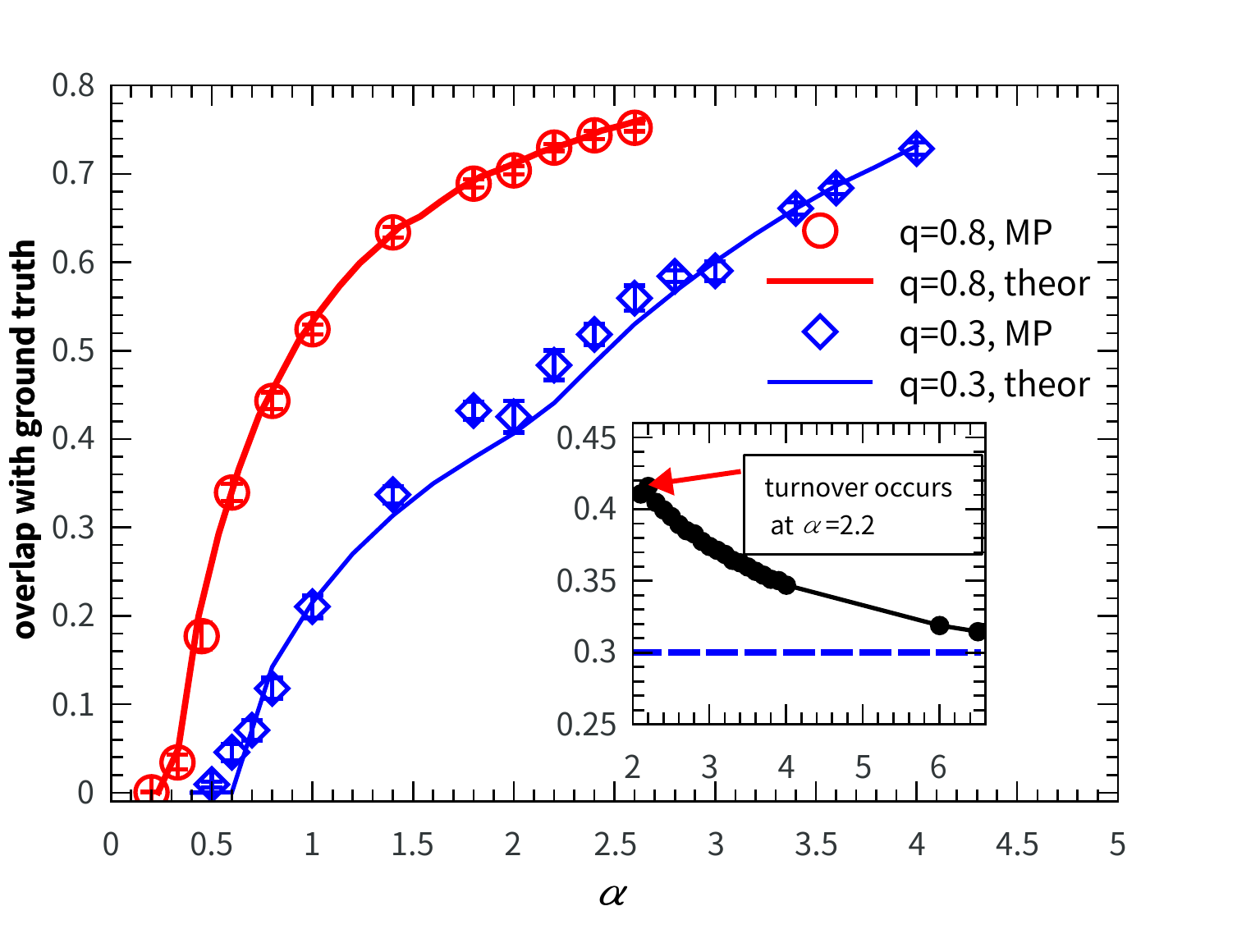}}
  \caption{
  (Color online) Learning performances of the minimal model obtained from message passing (MP) algorithmic results in comparison with theory. We consider $\beta=1$ with different values of $q$.
  MP is run on single instances of the minimal model with $N=200$. The error bars characterize the standard deviation across different random realizations of the minimal model.
  (a) Rescaled free energy per neuron as a function of data density (data samples per neuron).
   (b) The overlap with the ground truth versus the data density. In the inset, we show the replica prediction of the permutation-type overlap ($\tau_1$ or $\tau_2$) obtained 
   by an exchange of $\bx^1$ and $\bx^2$ in $T_1$ or $T_2$ for $q=0.3$, provided that $(\bx^{1,{\rm true}},\bx^{2,{\rm true}})$ is the planted feature. As expected, when the number of data samples is sufficiently large, this overlap tends to the embedded correlation $q$ (indicated by the dashed line).
  }\label{minmodel}
\end{figure}

First, we compare the free energy function estimated under the Bethe approximation with that predicted by the replica theory (in the thermodynamic limit).
As shown in Fig.~\ref{minmodel} (a), the algorithmic results on finite-sized networks coincide very well with the theoretical predictions. This implies that, the approximation we used to derive 
the message passing equation for learning a RBM with two hidden units is reasonable, especially when the number of visible neurons is large.

\begin{figure}
\centering
\subfigure[]{
     \includegraphics[bb=0 0 418 332,width=.46\textwidth]{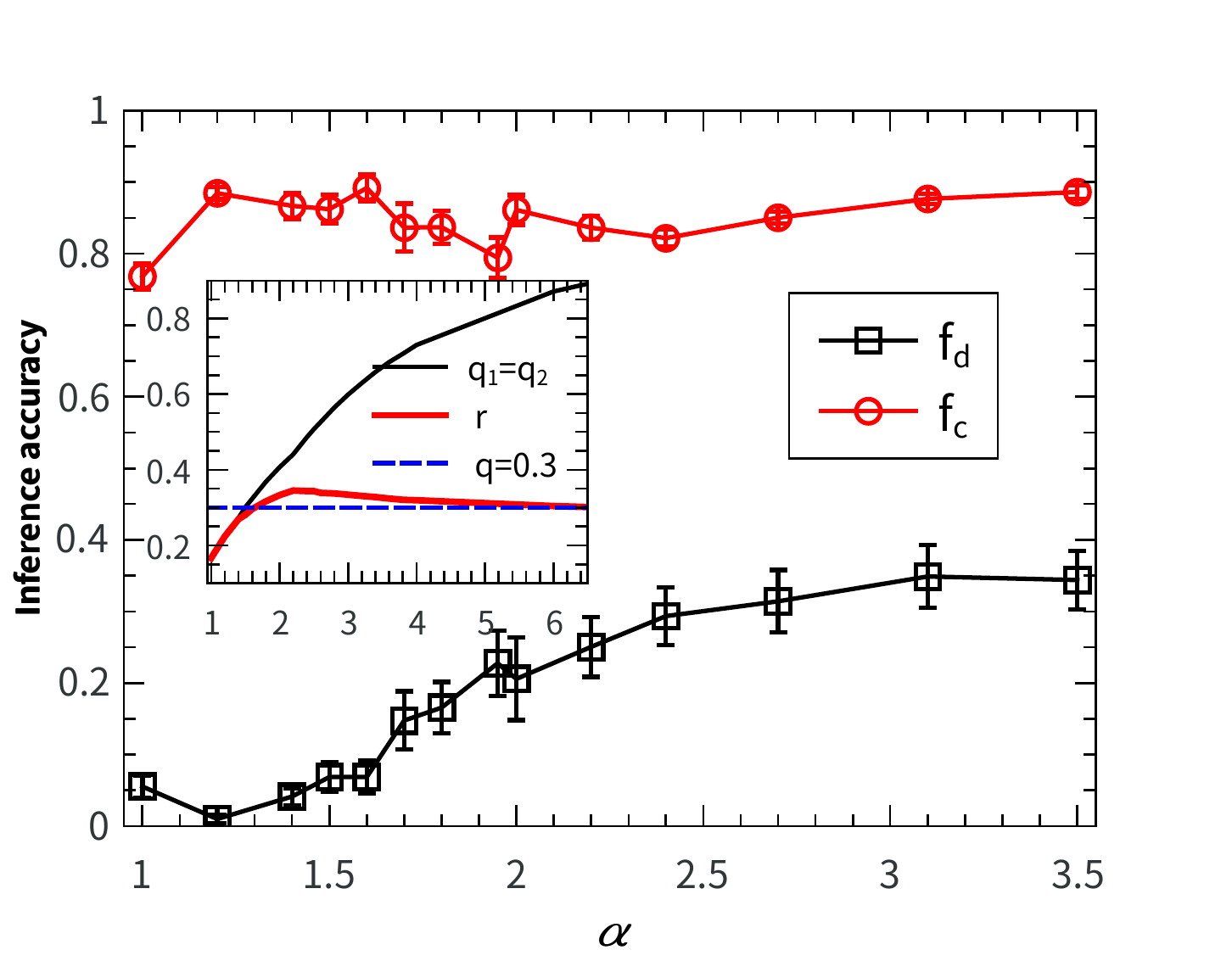}}
     \subfigure[]{\includegraphics[bb=0 0 412 334,width=.44\textwidth]{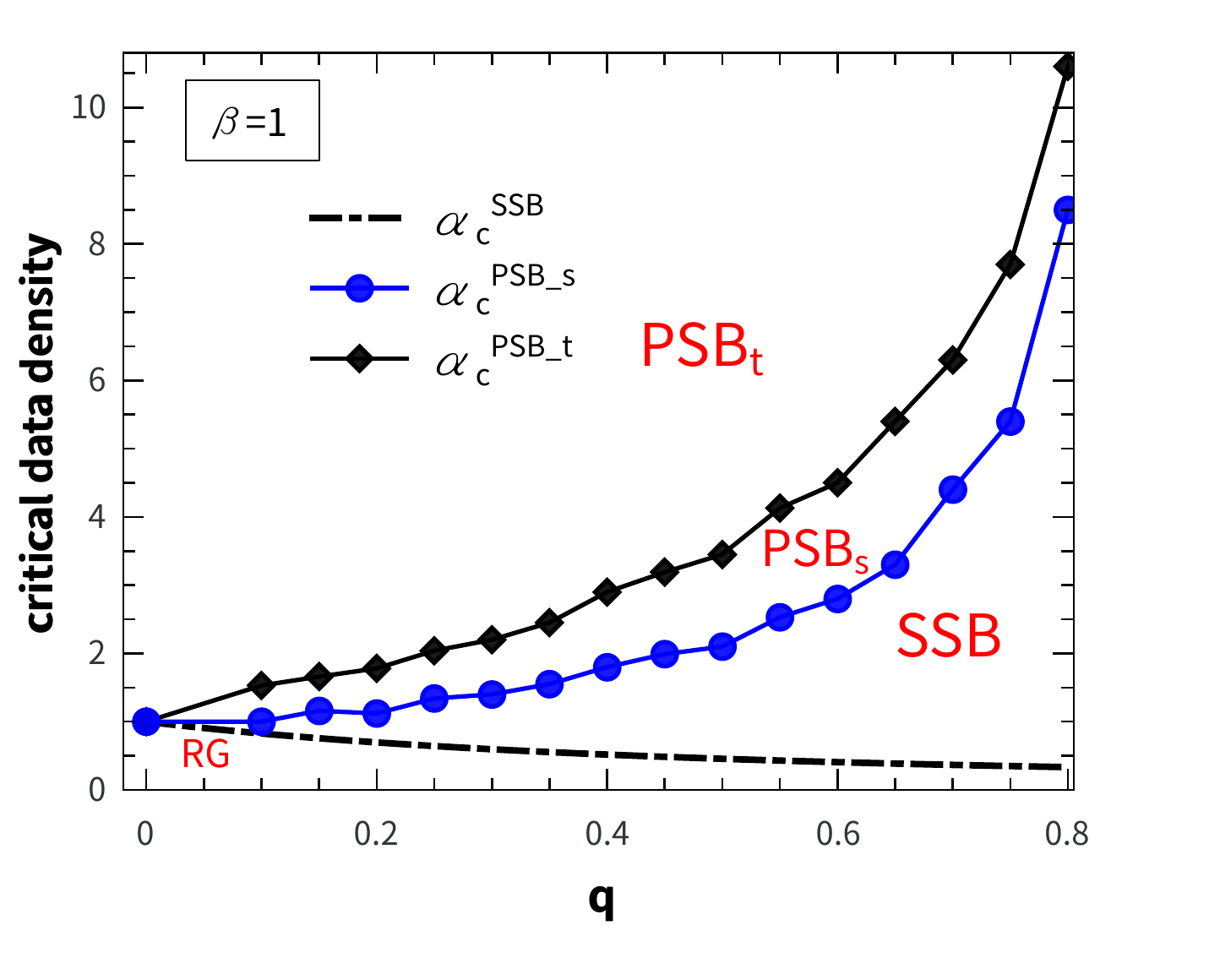}}
  \caption{
  (Color online) Phase diagram of unsupervised learning. 
  (a) Learning success probabilities on common feature components ($f_c$) and distinct components ($f_d$) show that the
  permutation symmetry can be spontaneously broken as predicted by the replica theory (the inset of Fig.~\ref{minmodel} (b)).
  Learning success of one component implies that the inferred value of that component matches with the true one.
  The inset shows the replica result of $q_1$ (or $q_2$) and $r$ versus the data density. 
  MP is run on single instances of the minimal model with $N=200$, $\beta=1$ and $q=0.3$.
  The error bars characterize the standard deviation across different random realizations of the minimal model.
   (b) Three phases (random-guess (RG), spontaneous symmetry breaking (SSB) and permutation symmetry breaking (PSB including ${\rm PSB_s}$ for inferred/student features and ${\rm PSB_t}$ for planted/teacher features))
   are separated by three curves--- $\alpha_c^{{\rm SSB}}(q)$, $\alpha_c^{{\rm PSB_s}}(q)$ and $\alpha_c^{{\rm PSB_t}}(q)$.
   $\alpha_c^{{\rm PSB_s}}$ can only be determined by numerically computing the deviation between $q_1$ (or $q_2$) and $r$, and $\alpha_c^{{\rm PSB_t}}$ is indicated by the point where $r$ starts to decrease (or the difference between $T_1$
   (or $T_2$) and $\tau_1$ (or $\tau_2$) starts to appear). We consider $\beta=1$ with different values of $q$.
  }\label{psb}
\end{figure}

Then, we study the evolution of the overlap with the ground truth as a function of the data density. As shown in Fig.~\ref{minmodel} (b), we clearly see a continuous phase transition separating a disordered (symmetric) phase from an ordered (symmetry-broken) phase.
This is consistent with previous works revealing the spontaneous symmetry breaking in unsupervised feature learning~\cite{Huang-2016b,Huang-2017}. When the amount of provided data samples is not too large,
the original model maintains its symmetry (i.e., equal probabilities for positive and negative assignments of synapses respectively). This is in an isotropic phase which does not capture any concept from the data samples. However, the increasing amount of data samples will 
break this symmetry through a continuous phase transition towards a non-trivial concept formation. During this process, in a practical message passing procedure, messages flowing in the factor graph are biased towards
the true feature maps underlying the noisy data. Remarkably,
the 
theory predicts the exact location of the phase transition point at which the messages running on a single instance start to polarize towards the true
feature map. As predicted and observed, the learning threshold indeed decreases as the absolute value of $q$ grows. For $|q|=0.3$, $\alpha_c\simeq0.596$; and for $|q|=0.8$, $\alpha_c$ is significantly reduced to be about $0.334$.

Due to the permutation symmetry, we find that after the SSB transition, there appear in sequence three non-trivial solutions to the saddle point equations (Eqs.~(\ref{rbmReplica2}) and~(\ref{rbmReplica3})), 
as $\alpha$ increases for a given value of $(\beta,q)$. The first solution has the form of $q_1=q_2=r$ and $T_1=T_2=\tau_1=\tau_2$.
This solution is caused by the permutation symmetry and is dominant at the earlier stage of the post-SSB-transition. $q_1=q_2=r$ implies the permutation symmetry between 
inferred vectors of $\bx^1$ and $\bx^2$; and $T_1=T_2=\tau_1=\tau_2$ implies the permutation symmetry between planted true feature vectors.
In other words, after the SSB transition, the unsupervised learning increases the overlap with the ground truth through
identifying the common components of the two true feature maps.

The SSB phase is stable until a point where the unsupervised learning starts to predict the different components (as shown in Fig.~\ref{psb} (a)), thereby breaking the 
symmetry between the inferred values of $\bx^1$ and $\bx^2$. This point is thus referred to as $\alpha_c^{{\rm PSB_s}}$, namely, the permutation symmetry breaking (PSB) for student/inferred features.
However, the inferred features are equally likely to be either $(\bx^{1,{\rm true}}, \bx^{2,{\rm true}})$ or its permutation, thus the permutation symmetry between planted/teacher features (i.e., $\bx^{1,{\rm true}}$ and $\bx^{2,{\rm true}}$) is still preserved, therefore, after the ${\rm PSB_s}$ transition, we have the second solution: $q_1=q_2\neq r$, and $T_1=T_2=\tau_1=\tau_2$.
This second solution is stable until there is a turnover of the trend of the permutation-type overlap or the order parameter $r$ (as shown in the insets of Fig.~\ref{minmodel} (b) and Fig.~\ref{psb} (a) respectively). At this turnover, $r$ starts to decrease,
thereby reducing the permutation-type overlap towards the true correlation level of the two feature maps (Fig.~\ref{minmodel} (b)).
After this transition indicated by the turnover, even the permutation symmetry between the teacher's features is broken, as the unsupervised learning is able to distinguish the two feature maps underlying the raw data samples.
Therefore, we observe the final solution that has the form of either $q_1=q_2=T_1=T_2$ or $q_1=q_2=\tau_1=\tau_2$ (two sub-forms), which is dominant at the later stage of the unsupervised learning. Accordingly, this second PSB transition
is called ${\rm PSB_t}$ transition with broken symmetry for teacher's features.

We remark that this third type non-trivial solution can be deduced from the fact that as $\alpha$ grows, the inferred feature map gets close to the true feature map
and thus both the inferred and true feature maps follow the same posterior probability of the learning process.
Note that for this solution, these two sub-forms share the same free energy. One inferred feature map has the freedom of choosing to match the first or second true feature map. This choice does not change the overall free energy.
Therefore, at the later stage, one solution of a
larger overlap corresponds to the case that the inferred feature maps are matched with their true counterparts; the other smaller overlap corresponds to the case that
the inferred feature maps are matched with their interchanged (permuted) counterparts. In particular, the latter one should converge to the embedded correlation $q$ as the learning converges to
the true feature maps. We show one example in the case of $q=0.3$ (see the inset of Fig.~\ref{minmodel} (b)), which is consistent with the replica prediction (the inset of Fig.~\ref{psb} (a)).
Note that these two kinds of overlaps are very hard to distinguish at the earlier stage, since the inferred feature maps are only partially correct compared with
their true counterparts. Indeed, in the algorithmic
simulations, we observe two solutions of different overlaps (one is larger than the other), which are very easy to distinguish at the later stage of the learning. 
In addition, considering the sign of each inferred (or true) feature vector, we have other multiple types of solutions, since a sign-reversed assignment of the feature map does not change the 
posterior probability (Eq.~(\ref{Pobs2})). For simplicity, we do not consider the sign-reversed case in the analysis.

Permutation symmetry of different types can be spontaneously broken. Both transitions are continuous.
We summarize these qualitatively different transitions in Fig.~\ref{psb} (b), showing that
as $q$ grows, the gap between $\alpha_c^{{\rm SSB}}$ and $\alpha_c^{{\rm PSB}}$ increases, thereby demonstrating that
the small-$q$ region is beneficial not only in the sense of a significant decrease of data size triggering the concept-formation but also in the sense of a small data size triggering the permutation symmetry breaking.
Therefore, not only can increasing the number of data samples (observations) drive a spontaneous symmetry breaking, but increasing the observations
also drive a permutation symmetry breaking which finally leads to a perfect reconstruction of the embedded feature maps.

In addition, when one tries to run the message passing
from a perturbed true
feature map, it is observed that its basin is quite large, i.e., it is stable in the presence of a sufficiently large dataset. But if the perturbation is large enough (e.g., crossing the sign boundary), the passing message is able to
converge to other types of fixed points (e.g., the sign-reversed or permutation-symmetric ones).


\section{Summary}
\label{sum}
In conclusion, we propose a minimal model of the permutation symmetry in a simple unsupervised learning system--- a simple RBM with only two hidden units. Using statistical mechanics tools developed in theory of
disordered systems, we reveal very rich properties of this model. Effects of permutation symmetry among hidden units have been studied 
in supervised learning in multilayer neural networks~\cite{Engel-1992,Barkai-1992}.
Here we are the first to consider its effects on unsupervised learning, focusing on spontaneous symmetry breaking and the critical data size that triggers the transition.

First, we derive an efficient algorithm to infer the embedded feature vectors from a given dataset (Monte-Carlo 
samples in this paper), according to the Bayesian principle. This algorithm is fully distributed, in that the computation is implemented
in terms of local passing messages among feature nodes and data (constraint) nodes. The algorithmic results can be used as a test of the replica
theory in the thermodynamic limit.

Second, the behavior of the algorithm, especially the critical data size at which a continuous phase transition from a disordered (symmetric) 
phase to an ordered (symmetry-broken) phase occurs, can be predicted by a replica theory of the minimal model. Due to the existence
of two hidden units, we have to manipulate eight order parameters that characterize all possible typical correlations in the
replica space itself as well as between replica space and true-feature space. In addition to the spontaneous symmetry breaking,
our theory predicts another later transition where the permutation symmetry is spontaneously broken and thus the identity of the feature map can be captured during unsupervised learning.
\textit{By studying the minimal model, we are able to interpret the unsupervised learning, a fundamental process governing artificial and biological intelligence, as a progressive combination of SSB and PSB (including two types---${\rm PSB_s}$ and ${\rm PSB_t}$),
both of which are driven by increasing the data size (or observations)}.

Based on the replica computation, we have three contributions. (i) We analytically prove that the critical data size 
for the phase transition indeed does not depend on the number of hidden units (a finite number case), and this conclusion was empirically observed in a previous work~\cite{Barra-2017}.
The critical value $\alpha_c=\beta^{-4}$ for learning in a one-bit RBM~\cite{Huang-2016b,Huang-2017} can directly apply to the RBM 
with two hidden units, once no correlations are embedded into the two feature vectors of hidden units. A detailed proof is also given in the~\ref{app4}. (ii) In addition, we reveal 
the correlation level of embedded true feature vectors reduces the critical data size, characterized by a simple formula (see Eq.~(\ref{thre})). As an example, in the very large $\beta$
limit (small variability in the data space), the critical data size for $q=-1$ is reduced 
to only one-fourth of that in the correlation-free case. In another limit $q\rightarrow0$, depending on the order of magnitude of $q$ compared with $\beta^{-2}$,
the necessary data size that triggers the transition can be reduced to one half of $\beta^{-4}$ or even less, although the receptive fields are weakly-correlated.
This prediction qualitatively coincides with the observation that
humans or non-human animals do not need a large amount of data samples to learn a concept from structured examples (e.g, natural images are highly structured
as various levels of correlations are embedded)~\cite{Lake-2017}. We expect that this quantitative prediction of
the critical learning data threshold can be shown to hold in a generic case with arbitrary levels of receptive-field-correlation, and with
an arbitrary finite number of hidden units, in particular for generative models of neural networks. (iii) Our theory predicts that an additional spontaneous permutation symmetry breaking follows
the spontaneous symmetry breaking, leading to another benefit of the small-$q$ regime where the data size triggering the permutation symmetry breaking is smaller compared with the large-$q$ regime (Fig.~\ref{psb}).

Our study also encourages several interesting future directions. First, by analogy with a Bayesian iteration derived in our 
previous work~\cite{Huang-2017} to predict the true noise level (the hyper-parameter $\beta$) in a dataset, one can
also derive the iteration equation for predicting both correlation level $q$ and noise level $\beta$ for an arbitrary dataset. Second, one can also 
verify the predicted value of $\alpha_c$ in a practical neural network architecture with inferred $q$ and $\beta$.
This can be carried out in a more complex RBM, by comparing the predicted value of the critical data size with the observed one for a successful unsupervised 
learning task. Furthermore, the prediction of our theory about 
the benefit of weakly-correlated feature maps compared with the correlation-free situation can also be tested in practical artificial neural networks and even in biological neural networks.
Finally, it would be very interesting to verify in a more general setting whether the progressive combination of SSB and PSB is the underlying mechanism of unsupervised learning.
In this sense, our minimal model paves the way towards understanding the fundamentally important unsupervised learning process, by addressing 
the role of the permutation symmetry and moreover 
the critical learning data sizes (related to both SSB and PSB) in a simple unsupervised learning system.

\section*{Acknowledgments}

We would like to thank referees for their inspiring comments. This research was supported by the start-up budget 74130-18831109 of the 100-talent-
program of Sun Yat-sen University (H.H.), the NSFC (Grant No. 11805284) (H.H.), and grants from the Research Grants Council 
of Hong Kong (16322616 and 16396817) (T.H. and M.W.)
\appendix
\section{Computation of $\left<\Omega^n\right>$}
\label{app1}

Here, we show details to compute  $\left<\Omega^{n}\right>$, 
which is defined as:
\begin{equation}
\label{partn}
    \begin{split}
         \langle  \Omega^{n} \rangle  = &\sum_{ \{ \bm{\xi }^{true},\bm{\sigma^{a}} \}  }  \prod_{i=1}^{N}[P(\xi_{i}^{1,true},\xi_{i}^{2,true} )] \prod_{a=1}^{M}  \frac{  \cosh{ ( \frac{\beta}{\sqrt{N}}\bm{\xi}^{1,true}\bm{\sigma}^{a} ) } \cosh{  (  \frac{\beta}{\sqrt{N}}   \bm{\xi}^{2,true}\bm{\sigma}^{a}   )  }}{2^{N}e^{\beta^{2}} \cosh{(\beta^{2}q})}\\
         &\times\sum_{ \{\bm{\xi}^{1,\gamma},\bm{\xi}^{2,\gamma}   \} } \prod_{a,\gamma}  \frac{\cosh{( \frac{\beta}{\sqrt{N}} \bm{\xi}^{1,\gamma}  \bm{\sigma}^{a} )  }   \cosh{( \frac{\beta}{\sqrt{N}} \bm{\xi}^{2,\gamma} \bm{\sigma}^{a}   } ) }{ \cosh{(\beta^{2}R^{\gamma}})  },
    \end{split}
\end{equation}
where $\gamma$ indicates the replica index, $\bm{\xi}^{true}= \{ \bm{\xi}^{1,true},\bm{\xi}^{2,true} \}  $, the overlap 
$ q=\frac{1}{N}\bm{\xi}^{1,true} \bm{\xi}^{2,true} $, and $ R^{\gamma}=\frac{1}{N}\bm{\xi}^{1,\gamma}  \bm{\xi}^{2,\gamma}   $.

To further calculate  $  \langle  \Omega^{n}   \rangle $, we need to define the order parameters as follows:
  \begin{subequations}
  \label{opdef}
    \begin{align}
        &T_{1}^{\gamma}=\frac{1}{N}\bm{\xi}^{1,true}\bm{\xi}^{1,\gamma},     &&    T_{2}^{\gamma}=\frac{1}{N}\bm{\xi}^{2,true}\bm{\xi}^{2,\gamma},\\
        &\tau_{1}^{\gamma}=\frac{1}{N}\bm{\xi}^{1,true}\bm{\xi}^{2,\gamma}, &&  \tau_{2}^{\gamma}=\frac{1}{N}\bm{\xi}^{2,true}\bm{\xi}^{1,\gamma}, \\
        &  q_{1}^{\gamma ,\gamma^{'}}=\frac{1}{N}\bm{\xi}^{1,\gamma}\bm{\xi}^{1,\gamma^{'}},  && q_{2}^{\gamma ,\gamma^{'}}=\frac{1}{N}\bm{\xi}^{2,\gamma}\bm{\xi}^{2,\gamma^{'}},\\
        &  R^{\gamma}=\frac{1}{N}\bm{\xi}^{1,\gamma}\bm{\xi}^{2,\gamma}, &&   r^{\gamma  ,\gamma^{'}}=\frac{1}{N} \bm{\xi}^{1,\gamma}\bm{\xi}^{2,\gamma^{'}}.
    \end{align}
  \end{subequations}
Note that these order parameters can be used to evaluate the disorder average in Eq.~(\ref{partn}). We then proceed as follows,
\begin{equation}
\label{partn2}
    \begin{split}
        &   \langle  \Omega^{n} \rangle=\sum_{\{ \bm{\sigma}^{a},\bm{\xi}^{true} \}}  \prod_{i=1}^{N}P(   \xi_{i}^{1,true},\xi_{i}^{2,true})  \sum_{ \{ \bm{\xi}^{1,\gamma},\bm{\xi}^{2,\gamma}  \}   }   \int   \prod_{\gamma=1}^{n} dR^{\gamma} \delta(\bm{\xi}^{1,\gamma} \bm{\xi}^{2,\gamma} -NR^{\gamma})  \\
        &\times\int  \prod_{\gamma=1}^{n} dT_{1}^{\gamma} \delta(\bm{\xi}^{1,true} \bm{\xi}^{1,\gamma} -NT_{1}^{\gamma})\int  \prod_{\gamma=1}^{n} dT_{2}^{\gamma}  \delta(\bm{\xi}^{2,true} \bm{\xi}^{2,\gamma} -NT_{2}^{\gamma}) \int \prod_{\gamma=1}^{n} d\tau_{1}^{\gamma}  \delta(\bm{\xi}^{1,true} \bm{\xi}^{2,\gamma} -N\tau_{1}^{\gamma})\\
        &\times  \int  \prod_{\gamma=1}^{n} d\tau_{2}^{\gamma}  \delta(\bm{\xi}^{2,true} \bm{\xi}^{1,\gamma} -N\tau_{2}^{\gamma}) \int \prod_{\gamma <\gamma^{'}}d q_{1}^{\gamma,\gamma^{'}}  \delta(\bm{\xi}^{1,\gamma} \bm{\xi}^{1,\gamma^{'}}-Nq_{1}^{\gamma,\gamma^{'} }) \int  \prod_{\gamma <\gamma^{'}}d q_{2}^{\gamma,\gamma^{'}}  \delta(\bm{\xi}^{2,\gamma} \bm{\xi}^{2,\gamma^{'}}-Nq_{2}^{\gamma,\gamma^{'} }) \\
        &  \times \int\prod_{\gamma <\gamma^{'}}d r^{\gamma,\gamma^{'}}  \delta(\bm{\xi}^{1,\gamma} \bm{\xi}^{2,\gamma^{'}}-Nr ^{\gamma,\gamma^{'} }) \prod_{a=1}^{M}\bigg \{   \frac{ \cosh{(\beta X^{0}_{a})}  \cosh{(\beta Y_{a}^{0})} }{ 2^{N}e^{\beta^{2}} \cosh{(\beta^{2}q)}} \prod_{\gamma=1}^{n} \frac{\cosh{(\beta X_{a}^{\gamma})}\cosh{(\beta Y_{a}^{\gamma})} }{   \cosh{(\beta^{2}R^{\gamma}  )}    }                    \bigg \}\\
        &=\sum_{\{ \bm{\sigma}^{a},\bm{\xi}^{true} \}} \prod_{i=1}^{N}P(\xi_{i}^{1,true},\xi_{i}^{2,true}  ) \sum_{  \{\bm{\xi}^{1,\gamma},\bm{\xi}^{2,\gamma} \}} \int \prod_{\gamma=1}^{n} \bigg(  \frac{ d R^{\gamma} d \hat{R}^{\gamma}  }{2\pi }  \bigg ) \int \prod_{\gamma=1}^{n} \bigg(  \frac{ d T_{1}^{\gamma} d \hat{T_{1}}^{\gamma}  }{2\pi }  \bigg )  \int \prod_{\gamma=1}^{n} \bigg(  \frac{ d T_{2}^{\gamma} d \hat{T_{2}}^{\gamma}  }{2\pi }  \bigg ) \\
        &\times   \int \prod_{\gamma=1}^{n} \bigg(  \frac{ d \tau_{1}^{\gamma} d \hat{\tau}_{1}^{\gamma}  }{2\pi }  \bigg )   \int \prod_{\gamma=1}^{n} \bigg(  \frac{ d \tau_{2}^{\gamma} d \hat{\tau}_{2}^{\gamma}  }{2\pi }  \bigg )
        \int \prod_{\gamma<\gamma^{'}} \bigg(  \frac{ d q_{1}^{\gamma,\gamma^{'}}  d\hat{q}_{1}^{\gamma,\gamma^{'}}      }{2\pi}  \bigg) \int \prod_{\gamma<\gamma^{'}} \bigg(  \frac{ d q_{2}^{\gamma,\gamma^{'}}  d\hat{q}_{2}^{\gamma,\gamma^{'}}d r^{\gamma,\gamma^{'}}  d\hat{r}^{\gamma,\gamma^{'}} }{4\pi^2}  \bigg)  \\
        &\times  \exp \bigg (  \sum_{\gamma=1}^{n} i\hat{R}^{\gamma}(\bm{\xi}^{1,\gamma}\bm{\xi}^{2,\gamma}-NR^{\gamma}) +\sum_{\gamma=1}^{n} i\hat{T_{1}}^{\gamma}(\bm{\xi}^{1,\gamma}\bm{\xi}^{1,true}-NT_{1}^{\gamma}) +\sum_{\gamma=1}^{n} i\hat{T_{2}}^{\gamma}(\bm{\xi}^{2,\gamma}\bm{\xi}^{2,true}-NT_{2}^{\gamma}) ) \bigg )  \\
        & \times  \exp \bigg( \sum_{\gamma=1}^{n}i\hat{\tau_{1}}^{\gamma}(\bm{\xi}^{1,true}\bm{\xi}^{2,\gamma}-N\tau_{1}^{\gamma})  +   \sum_{\gamma=1}^{n}i\hat{\tau_{2}}^{\gamma}(\bm{\xi}^{2,true}\bm{\xi}^{1,\gamma}-N\tau_{2}^{\gamma} ) +\sum_{\gamma<\gamma^{'}}\hat{q_{1}}^{\gamma,\gamma^{'}}( \bm{\xi}^{1,\gamma}\bm{\xi}^{1,\gamma'}-Nq_{1}^{\gamma,\gamma^{'}}  )   \bigg )\\
        &  \times \exp  \bigg(  \sum_{\gamma<\gamma^{'}}\hat{q_{2}}^{\gamma,\gamma^{'}}( \bm{\xi}^{2,\gamma}\bm{\xi}^{2,\gamma^{'}}-Nq_{2}^{\gamma,\gamma^{'}}  ) +  \sum_{\gamma<\gamma^{'}}\hat{r}^{\gamma,\gamma^{'}}( \bm{\xi}^{1,\gamma}\bm{\xi}^{2,\gamma^{'}}-Nr^{\gamma,\gamma^{'}}  ) \bigg )\\
        &\times\prod_{a=1}^{M}\bigg \{   \frac{ \cosh{(\beta X^{0}_{a})}  \cosh{(\beta Y_{a}^{0})} }{ 2^{N}e^{\beta^{2}} \cosh{(\beta^{2}q)}}     
          \prod_{\gamma=1}^{n} \frac{\cosh{(\beta X_{a}^{\gamma})}\cosh{(\beta Y_{a}^{\gamma})} }{   \cosh{(\beta^{2}R^{\gamma}  )}    }  \bigg \},
    \end{split}
\end{equation}
where we have defined  $X^{ 0 }_{a} =\frac{1}{\sqrt{N}}\sum_{i=1}^{N}\xi_{i}^{1,true}\sigma_{i}^{a}$, $Y^{ 0 }_{a} =\frac{1}{\sqrt{N}}\sum_{i=1}^{N}\xi_{i}^{2,true}\sigma_{i}^{a} $, and $X_a^{ \gamma } =\frac{1}{\sqrt{N}}\sum_{i=1}^{N}\xi_{i}^{1,\gamma}\sigma_{i}^{a}$, $ Y^{ \gamma }_{a} =\frac{1}{\sqrt{N}}\sum_{i=1}^{N}\xi_{i}^{2,\gamma}\sigma_{i}^{a} $.
To get the second equality, we have used the integral representation of the delta function $\delta(x)=\int \frac{d \hat{x}}{2\pi} e^{i \hat{x}x  }$.

To compute the free energy value, we assume a simple ansatz, i.e.,  all order parameters do not depend on their specific replica indexes,  which  is called  the replica-symmetry assumption. To be more precise, we assume
\begin{subequations}
\label{RS1}
  \begin{align}
      &  R^{\gamma}=R,  &&  i\hat{R} ^{\gamma}=\hat{R},  \\
      &  T_{1}^{\gamma}=T_{1},  &&  i\hat{T_{1}}^{\gamma}=\hat{T_{1}}, \\
      &  T_{2}^{\gamma}=T_{2},  &&  i\hat{T_{2}}^{\gamma}=\hat{T_{2}},\\
      &  \tau_{1}^{\gamma}=\tau_{1},  &&  i\hat{\tau}_{1}^{\gamma}=\hat{\tau_{1}}, \\
      &  \tau_{2}^{\gamma}=\tau_{2},  &&  i\hat{\tau}_{2}^{\gamma}=\hat{\tau_{2}},
  \end{align}
\end{subequations}
for any $\gamma$. We also assume that
\begin{subequations}
\label{RS2}
  \begin{align}
      &   q_{1}^{\gamma,\gamma^{'}}=q_{1},  &&    i\hat{q_{1}}^{\gamma,\gamma^{'}}=\hat{q_{1}},  \\
      &   q_{2}^{\gamma,\gamma^{'}}=q_{2},  &&    i\hat{q_{2}}^{\gamma,\gamma^{'}}=\hat{q_{2}},   \\
      &   r^{\gamma,\gamma^{'}}=r,    &&   i\hat{r}^{\gamma,\gamma^{'}}=\hat{r}, 
  \end{align}
\end{subequations}
for any  $\gamma$ and $\gamma^{'}$. Then we can express $ \langle  \Omega^{n}     \rangle   $ as :
\begin{equation}
\label{sdm}
    \langle   \Omega^{n}  \rangle=\int d \mathcal{O} d\mathcal{\Hat{O}}  e^{N  \mathcal{A} (\mathcal{O},\hat{\mathcal{O}},\alpha ,\beta,n) }.
\end{equation}
In the thermodynamics limit, $\langle \Omega^{n}  \rangle   $ can be approximated as $e^{N \mathcal{A}(\mathcal{O}_*,\hat{\mathcal{O}}_*,\alpha ,\beta,n) } $(namely, the saddle-point method), where $\mathcal{O}_*$  and  $\mathcal{\hat{O}}_*$ represent all non-conjugated order parameters and
conjugated order parameters evaluated at the maximal value of the action, respectively.
The expression for the action $\mathcal{A} (\mathcal{O},\hat{\mathcal{O}},\alpha ,\beta,n)  $ 
can be written by 
 \begin{equation}
 \label{sdm2}
 \begin{split}
     \mathcal{A}&=- nR\hat{R}-nT_{1}\hat{T_{1}}-nT_{2}\hat{T_{2}}-n\tau_{1}\hat{\tau_{1}}-n\tau_{2}\hat{\tau_{2}}-\frac{n(n-1)}{2}q_{1}\hat{q_{1}}\\
     &-\frac{n(n-1)}{2}q_{2}\hat{q_{2}}-\frac{n(n-1)}{2}r\hat{r}+G_{S}+\alpha G_{E},
     \end{split}
 \end{equation}
where $G_{S}$ is the entropy term, and $G_{E}$ is the energy term.

To derive the entropy term $G_{S}$, we use the following identities\cite{Huang-2013}:
\begin{subequations}
\label{gseq}
  \begin{align}
      & \sum_{\gamma <\gamma^{'}}\xi^{1,\gamma}\xi^{1,\gamma^{'}}=\frac{1}{2} \left(\sum_{\gamma} \xi^{1,\gamma} \right)^{2}- \frac{1}{2} \sum_{\gamma}(\xi^{1,\gamma})^{2},  \\
      & \sum_{\gamma <\gamma^{'}}\xi^{2,\gamma}\xi^{2,\gamma^{'}}=\frac{1}{2} \left(\sum_{\gamma} \xi^{2,\gamma} \right)^{2}- \frac{1}{2} \sum_{\gamma}(\xi^{2,\gamma})^{2},  \\
      \begin{split}
        & \sum_{\gamma <\gamma^{'}}\xi^{1,\gamma}\xi^{2,\gamma^{'}}= \frac{1}{2} \sum_{\gamma,
        \gamma^{'}}\xi^{1,\gamma}\xi^{2,\gamma^{'}}-\frac{1}{2}\sum_{\gamma}\xi^{1,\gamma}\xi^{2,\gamma} \\
      & = \frac{1}{4}\left( \sum_{\gamma}\xi^{1,\gamma}+\sum_{\gamma^{'}}\xi^{2,\gamma'}\right)^{2}-\frac{1}{4}\left(\sum_{\gamma} \xi^{1,\gamma}\right )^{2}-\frac{1}{4}\left(\sum_{\gamma^{'}} \xi^{2,\gamma'} \right)^{2}-\frac{1}{2}\sum_{\gamma}\xi^{1,\gamma}\xi^{2,\gamma}.
     \end{split}    
    \end{align}
 \end{subequations}
The above non-linear terms can be reduced to linear terms in 
the exponential functions of Eq.~(\ref{sdm}) by the Hubbard-Stratonovich  
transformation  $ \int Dt e^{bt}=e^{\frac{1}{2}b^{2}}$. 
Then, we obtain $G_{S}$ as :
\begin{equation}
\label{gs}
    \begin{split}
        G_{S}&= \ln\left[ \sum_{ \{ \xi^{1,\gamma}\xi^{2,\gamma}  \}}
        \exp\left( \hat{R}\sum_{\gamma=1}^{n}\xi^{1,\gamma}\xi^{2,\gamma}+
        \hat{T_{1}}\sum_{\gamma=1}^n \xi^{1,\gamma}\xi^{1,true}+
        \hat{T_{2}}\sum_{\gamma=1}^{n}\xi^{2,\gamma}\xi^{2,true}+
        \hat{\tau_{1}}\sum_{\gamma=1}^n\xi^{1,true}\xi^{2,\gamma}\right)\right.\\
        &\left.\times\exp\left(\hat{\tau_{2}}\sum_{\gamma=1}^{n} \xi^{1,\gamma}\xi^{2,true}+ 
        \hat{q_{1}}\sum_{\gamma<\gamma^{'}} \xi^{1,\gamma}\xi^{1,\gamma^{'}}
             +\hat{q_{2}}\sum_{\gamma<\gamma^{'}} \xi^{2,\gamma}\xi^{2,\gamma^{'}}
             +\hat{r}\sum_{\gamma<\gamma^{'}} \xi^{1,\gamma}\xi^{2,\gamma^{'}}\right)\right]_{\xi^{1,true}, \xi^{2,true}}\\
             &= \ln\left[\sum_{\{\xi^{1,\gamma},\xi^{2,\gamma}\}} \exp\left(\frac{(\hat{q_{1}}- \frac{\hat{r}  }{2}  )   }{2} \left(\sum_{\gamma} \xi^{1,\gamma}\right)^{2} 
             + \frac{ (\hat{q_{2}}- \frac{\hat{r}  }{2}  )  }{2}  \left(\sum_{\gamma} \xi^{2,\gamma}\right)^{2}
             + \hat{T_{1}}\sum_{\gamma}\xi^{1,\gamma}\xi^{1,true}
             \right)\right.\\               \\
        &\left.\times\exp\left( \frac{\hat{r}}{4}\left(\sum_{\gamma}\xi^{1,\gamma}       
        +\sum_{\gamma^{'}}\xi^{2,\gamma^{'}} \right)^{2}  
        +\hat{T_{2}}\sum_{\gamma}\xi^{2,\gamma}\xi^{2,true}+(\hat{R}-\frac{\hat{r}}{2})\sum_{\gamma}\xi^{1,\gamma}\xi^{2,\gamma}\right)\right.\\
        &\left.\times\exp\left( \hat{\tau_{1}}\sum_{\gamma}\xi^{1,true}\xi^{2,\gamma}+
        \hat{\tau_{2}}\sum_{\gamma}\xi^{2,true}\xi^{1,\gamma}-\frac{n}{2}\hat{q}_{1}-\frac{n}{2}\hat{q_{2}}\right)\right]_{\xi^{1,true},\xi^{2,true}}\\
        &=   \ln \left[ \sum_{ \{ \xi^{1,\gamma},\xi^{2,\gamma}  \} } \int D\mathbf{z}\exp\left( \sum_{\gamma}\sqrt{\hat{q_{1}}-\frac{\hat{r}}{2}}\xi^{1,\gamma}z_{1} 
        +\sum_{\gamma}\sqrt{\hat{q_{2}}-\frac{\hat{r}}{2}}\xi^{2,\gamma}z_{2}
        +\sqrt{\frac{\hat{r}}{2}}z_{3}\left(\sum_{\gamma}\xi^{1,\gamma}+\sum_{\gamma^{'}}\xi^{2,\gamma^{'}}\right)\right)\right.  \\
        & \left.\times\exp\left(\hat{T_{1}}\sum_{\gamma}\xi^{1,true}\xi^{1,\gamma}+\hat{T_{2}}\sum_{\gamma}\xi^{2,\gamma}\xi^{2,true}+ 
        \hat{\tau_{1}}\sum_{\gamma}\xi^{1,true}\xi^{2,\gamma}\right)\right.\\
        &\left.\times\exp\left(\hat{\tau_{2}}\sum_{\gamma}\xi^{2,true}\xi^{1,\gamma}
        +(\hat{R}-\frac{\hat{r}}{2})\sum_{\gamma}\xi^{1,\gamma}\xi^{2,\gamma}-\frac{n}{2}\hat{q}_{1}-\frac{n}{2}\hat{q_{2}}\right)\right]_{\xi^{1,true},\xi^{2,true}}.
    \end{split}
\end{equation}
Finally, we can express the entropy term $G_{S}$ in a compact form as
\begin{equation}
\label{gscomp}
    G_{S}= \ln\left[  \int D\mathbf{z}\left ( \sum_{\xi^{1},\xi^{2}}  e^{  b_{1}\xi^{1}+b_{2}\xi^{2}+b_{3}\xi^{1}\xi^{2}   }\right)^{n} \right]_{\xi^{1,true},\xi^{2,true}}-\frac{n}{2}\hat{q}_{1}-\frac{n}{2}\hat{q}_{2},
\end{equation}
where we have defined $D\mathbf{z}=Dz_{1}Dz_{2}Dz_{3} $, and the auxiliary variables $b_{1},b_{2},$ and $b_{3} $ as 
\begin{subequations}
\label{b123}
  \begin{align}
      &  b_{1}=\sqrt{\hat{q_{1}}-\frac{\hat{r}}{2} }z_{1}+\sqrt{\frac{\hat{r}}{2} }z_{3}    +\hat{T_{1}}\xi^{1,true}+\hat{\tau_{2}}\xi^{2,true},\\
      &   b_{2}=\sqrt{ \hat{q_{2}}-\frac{\hat{r}}{2} }z_{2}+\sqrt{\frac{\hat{r}}{2} }z_{3}+\hat{T_{2}}\xi^{2,true}+\hat{\tau_{1}}\xi^{1,true},\\
      & b_{3}=\hat{R}-\frac{\hat{r}}{2}.
  \end{align}
\end{subequations}
We remark that in the expression of $G_{S}$, the inner summation over $\xi^{1},\xi^{2}$ can be thought as a two-spin interaction partition function, which is defined as $Z_{{\rm eff}}$ in the main text. 
$ [ \bullet ]_{\xi^{1,true},\xi^{2,true} }  $ means an average w.r.t $ P(\xi^{1,true},\xi^{2,true}) $ which is also defined in the main text.

Next, we turn to compute the energy term $ G_{E} $. The expression of $G_{E}$ is given by
\begin{equation}
\label{gedef}
    G_{E}=\ln \left< \frac{ \cosh{(\beta X^{0})}\cosh{(\beta Y^{0}})     }{\cosh{(\beta^{2}q})}   \prod_{\gamma=1}^{n}  \frac{ \cosh{(\beta X^{\gamma})}\cosh{(\beta Y^{\gamma}} )}{\cosh{(\beta^{2}R^{\gamma})}} \right>, 
\end{equation}
 where $\langle   \bullet \rangle  $ defines the disorder average. 
$ X^{0}$, $Y^{0}$, $X^{\gamma}$, $Y^{\gamma}$ are correlated Gaussian random variables, which are the same as before but data index $a$ has been dropped off. They have zero mean and unit variance.
Their covariances are determined 
by the aforementioned order parameters as follows:
  \begin{subequations}
  \label{covarjust}
  \label{covar}
    \begin{align}
        & \langle  X^{0}Y^{0}   \rangle=q,  &&\langle  X^{0}X^{\gamma}  \rangle=T_{1},    &&    \langle X^{0}Y^{\gamma}  \rangle=\tau_{1},\\
        &\langle  X^{\gamma}X^{\gamma^{'}}    \rangle=q_{1},   && \langle  Y^{\gamma}Y^{\gamma^{'}}    \rangle=q_{2},    && \langle   X^{\gamma}Y^{\gamma}    \rangle=R, \\
        &  \langle  Y^{0}Y^{\gamma}  \rangle=T_{2},  && \langle  Y^{0}X^{\gamma}  \rangle=\tau_{2},    &&  \langle  X^{\gamma}Y^{\gamma^{'}} \rangle=r.
    \end{align}
  \end{subequations}
The random variables $X^{0},Y^{0},X^{\gamma},Y^{\gamma} $ can thus be parameterized by six standard Gaussian variables of zero mean and unit variance ($ t_{0},x_{0},u,u^{'},y_{\gamma},\omega_{\gamma} $) as follows,
\begin{subequations}
\label{para}
  \begin{align}
         X^{0}&=t_{0},\\
        Y^{0}&=qt_{0}+\sqrt{1-q^{2}}x_{0},\\
        X^{\gamma}&=T_{1}t_{0}+\frac{\tau_{2}-T_{1}q}{\sqrt{1-q^{2}}}x_{0}+Bu+\sqrt{1-q_{1}}\omega_{\gamma}, \\
      \begin{split}
           Y^{\gamma} &=\tau_{1}t_{0}+\frac{T_{2}-\tau_{1}q}{\sqrt{1-q^{2}}}x_{0}+\frac{r-A}{B}u+\frac{R-r}{\sqrt{1-q_{1}}}\omega_{\gamma}+Ku^{'} \\
           &  +\sqrt{1-q_{2}- \frac{(R-r)^{2}}{1-q_{1}}}y_{\gamma},
      \end{split} 
  \end{align}
\end{subequations}
where $A=T_{1}\tau_{1}+\frac{(\tau_{2}-T_{1}q)(T_{2}-\tau_{1}q)}{1-q^{2}}  $, 
$ B=\sqrt{ q_{1}-(T_{1})^{2}-\frac{(\tau_{2}-T_{1}q)^{2} }{ 1-q^{2}   }          }   $, and $  K=\sqrt{ q_{2}-(\tau_{1})^{2}-\frac{(T_{2}-\tau_{1}q)^{2}}{1-q^{2}}-(\frac{r-A}{B})^{2}      }$.
One can easily verify that the above parameterization satisfies their covariance structures. Therefore, the $G_{E}$ term can be calculated by a standard Gaussian integration given by
\begin{equation}
\label{GEres}
    \begin{split}
        & G_{E}=\ln   \left [ \int Dt_{0}Dx_{0} Du Du^{'}  \frac{ \cosh{(\beta t^{0})}   \cosh{\beta(qt_{0}+\sqrt{1-q^{2}}x_{0})} }{\cosh{(\beta^{2}q})}\right.\\
        &\left.\times \left( \int D \omega  Dy
         \frac{1}{\cosh{(\beta^{2}R)}}
          \cosh{\beta(T_{1}t_{0}+\frac{\tau_{2}-T_{1}q}{\sqrt{1-q^{2}}}x_{0}+Bu+\sqrt{1-q_{1}}\omega           )}\right.\right.\\
         & \left.\left.\times\cosh\beta(\tau_{1}t_{0}+\frac{T_{2}-\tau_{1}q}{\sqrt{1-q^{2}}}x_{0}+ \frac{r-A}{B} u+ +\frac{R-r}{\sqrt{1-q_{1}}}\omega+Ku^{'}+Cy ) \right)^{n} \right ],
        \end{split}
\end{equation}
where $C\equiv\sqrt{1-q_2-\frac{(R-r)^2}{1-q_1}}$.
To proceed, we first define the auxiliary quantities as :
\begin{subequations}
\label{Lam}
  \begin{align}
      &  \Lambda_{+}=(T_{1}+\tau_{1})t_{0}+\frac{ (T_{2}+\tau_{2})- q(T_{1}+\tau_{1}) }{ \sqrt{1-q^{2}} }x_{0}+(B+\frac{r-A}{B})u+Ku^{'}, \\
      &  \Lambda_{-}=(T_{1}-\tau_{1})t_{0}+\frac{ (\tau_{2}-T_{2})- q(T_{1}-\tau_{1}) }{ \sqrt{1-q^{2}} }x_{0}+(B-\frac{r-A}{B})u-Ku^{'}.
  \end{align}
\end{subequations}
Then we compute the integral inside the power $n$ defined by $I$ whose result is given by
\begin{equation}
\label{gecomput}
    \begin{split}
        & I\equiv\int D\omega Dy \bigg[ \cosh{\beta(\tau_{1}t_{0}+\frac{T_{2}-\tau_{1}q}{\sqrt{1-q^{2}}}x_{0}+\frac{r-A}{B}u+\frac{R-r}{\sqrt{1-q_{1}}}\omega+Ku^{'}+\sqrt{1-q_{2}- \frac{(R-r)^{2}}{1-q_{1}}}y)}\\
        & \times\cosh{\beta(T_{1}t_{0}+\frac{\tau_{2}-T_{1}q}{\sqrt{1-q^{2}}}x_{0}+Bu+\sqrt{1-q_{1}}\omega )}\bigg ]\\
        &=\frac{1}{4}\int  D\omega Dy  \bigg[  e^{ \beta\{\Lambda_{+}+(      \sqrt{1-q_{1}}+\frac{R-r}{\sqrt{1-q_{1}}})\omega+\sqrt{1-q_{2}-\frac{(R-r)^{2}}{1-q_{1}}}y \}}  
        + e^{ -\beta\{\Lambda_{+}+(      \sqrt{1-q_{1}}+\frac{R-r}{\sqrt{1-q_{1}}})\omega+\sqrt{1-q_{2}-\frac{(R-r)^{2}}{1-q_{1}}}y \}}\\
        &+e^{ \beta\{\Lambda_{-}+(      \sqrt{1-q_{1}}-\frac{R-r}{\sqrt{1-q_{1}}})\omega-\sqrt{1-q_{2}-\frac{(R-r)^{2}}{1-q_{1}}}y \}}+
         e^{ -\beta\{\Lambda_{-}+(      \sqrt{1-q_{1}}-\frac{R-r}{\sqrt{1-q_{1}}})\omega-\sqrt{1-q_{2}-\frac{(R-r)^{2}}{1-q_{1}}}y \}} \bigg ]\\
         &=\frac{1}{2}e^{\beta^{2} ( 1- \frac{q_{1}+q_{2}}{2})}  \bigg [e^{\beta^{2}(R-r)}\cosh{(\beta  \Lambda_{+})} +  e^{-\beta^{2}(R-r)}\cosh{(\beta  \Lambda_{-})}   \bigg].
    \end{split}
\end{equation}
For simplicity, we also define the following auxiliary quantities $ Z_{E},G_{c}^{-},G_{s}^{+},G_{s}^{-}$:

\begin{equation}
\label{AQ}
    \begin{split}
        &  Z_{E}= e^{\beta^{2}(R-r)}\cosh{(\beta  \Lambda_{+})} +  e^{-\beta^{2}(R-r)}\cosh{(\beta  \Lambda_{-})},  \\
        & G_{c}^{-}= \frac{ e^{\beta^{2}(R-r)}\cosh{(\beta  \Lambda_{+})} - e^{-\beta^{2}(R-r)}\cosh{(\beta  \Lambda_{-})}                           }{ e^{\beta^{2}(R-r)}\cosh{(\beta  \Lambda_{+})} +  e^{-\beta^{2}(R-r)}\cosh{(\beta  \Lambda_{-}} )},\\
        &G_{s}^{+}=\frac{ e^{\beta^{2}(R-r)}\sinh{(\beta  \Lambda_{+})} +  e^{-\beta^{2}(R-r)}\sinh{(\beta  \Lambda_{-})}       }{e^{\beta^{2}(R-r)}\cosh{(\beta  \Lambda_{+})} +  e^{-\beta^{2}(R-r)}\cosh{(\beta  \Lambda_{-})}},\\
        &G_{s}^{-}=\frac{ e^{\beta^{2}(R-r)}\sinh{(\beta  \Lambda_{+})} -  e^{-\beta^{2}(R-r)}\sinh{(\beta  \Lambda_{-})}    }{e^{\beta^{2}(R-r)}\cosh{(\beta  \Lambda_{+})} +  e^{-\beta^{2}(R-r)}\cosh{(\beta  \Lambda_{-})}}.
    \end{split}
\end{equation}
Following the replica trick, we can get:
\begin{equation}
\label{repge}
    \lim_{n \to 0} \frac{ G_{E} }{ n}=\frac{ \int Dt_{0}Dx_{0}DuDu^{'} \frac{\cosh{\beta  t_{0} } \cosh{\beta (qt_{0}+\sqrt{1-q^{2}}x_{0}    )}  }{  \cosh{\beta^{2}q}}   \ln\left[ \frac{I}{\cosh\beta^2R}\right]}{\int Dt_{0}Dx_{0}DuDu^{'}  \frac{\cosh{(\beta t_{0}) } \cosh\beta  (qt_{0}+\sqrt{1-q^{2}}x_{0})}{  \cosh{(\beta^{2}q})}    },
\end{equation}
where the integral in the denominator can be exactly computed with the result given by
\begin{equation}
\label{denom}
    \begin{split}
        &\int Dt_{0}Dx_{0}DuDu^{'} \cosh{(\beta t_{0}) } \cosh\beta  (qt_{0}+\sqrt{1-q^{2}}x_{0})\\
        &=\frac{1}{2}\left(e^{\frac{\beta^{2}}{2}(1-q)^{2}+\frac{\beta^{2}}{2}(1-q^{2}) }+   e^{\frac{\beta^{2}}{2}(1+q)^{2}+\frac{\beta^{2}}{2}(1-q^{2}) }  \right )=e^{\beta^{2}}\cosh{\beta^{2}q}.
    \end{split}
\end{equation}

Finally, by collecting all the above relevant terms,
we have the following estimation of $\left<\Omega^n\right>$ given by
\begin{equation}
\label{Znres}
    \begin{split}
    \langle \Omega^{n}\rangle=& \int d \mathcal{O}d\mathcal{\hat{O}} \exp\left(-NnR\Hat{R}-NnT_{1}\Hat{T_{1}}-NnT_{2}\Hat{T_{2}}-Nn\tau_{2}\Hat{\tau_{2}}-\frac{N}{2}n(n-1)q_{1}\Hat{q_{1}}\right)\\
    &\times\exp\left(-\frac{N}{2}n(n-1)q_{2}\Hat{q_{2}}-\frac{N}{2}n(n-1)r\hat{r}-\frac{nN}{2}\hat{q_{1}}-\frac{nN}{2}\hat{q_{2}} 
    +N\ln \left[\int D\mathbf{z} Z_{{\rm eff}}^{n}\right]_{\xi^{1,true},\xi^{2,true}}\right.\\
    &\left.+\alpha N\ln\left\{ \int D\mathbf{t}  \frac{ \cosh{(\beta t_{0}) } \cosh\beta  (qt_{0}+\sqrt{1-q^{2}}x_{0}) }{\cosh{(\beta^{2}q})} \left[\frac{I }{\cosh(\beta^{2}R)}\right]^{n}\right\}\right),
    \end{split}
\end{equation}
where in shorthand $D\mathbf{t}=Dt_0Dx_0DuDu'$.
By computing $\lim_{n \to 0}  \frac{\ln{\langle \Omega^{n}  \rangle }}{n} $ and using Eq.~(\ref{repge}), we get the expression $F_{\beta}= -\beta f_{RS}$ as

\begin{equation}
\label{freeErep}
    \begin{split}  
    & F_{\beta}= -R\hat{R}-T_{1}\hat{T_{1}}-T_{2}\hat{T_{2}}-\tau_{1}\hat{\tau_{1}}-\tau_{2}\hat{\tau_{2}}+\frac{\hat{q_{1}}}{2}(q_{1}-1)+\frac{\hat{q_{2}}}{2}(q_{2}-1)\\
    &+\frac{r\hat{r}}{2}+\int D\mathbf{z}\left[ \ln Z_{{\rm eff}} \right]_{\xi^{1,true},\xi^{2,true}}
    -\alpha\ln\left(2\cosh(\beta^{2}R)\right)+\alpha \beta^{2}\left(1-\frac{q_{1}+q_{2}}{2}\right)\\
    &+ \frac{  \alpha e^{-\beta^{2}}}{\cosh{(\beta^{2}q})} \int D\mathbf{t}\cosh{\beta t_{0}} \cosh{\beta (qt_{0}+\sqrt{1-q^{2}}x_{0})} \ln Z_{E}.
    \end{split}
\end{equation}
Note that we have used $ \lim_{n\to 0}  \frac{\ln \left[{\int D\mathbf{z} Z^{n}_{{\rm eff}} }\right]_{\xi^{1,true},\xi^{2,true}}}{n}
=\int D\mathbf{z}  \left[\ln Z_{{\rm eff}}\right]_{\xi^{1,true},\xi^{2,true}} $
to arrive at the final expression.

\section{Derivation of saddle-point equations}
\label{app2}
By the saddle-point analysis, these non-conjugated order parameters $\mathcal{O} $ should obey the following stationary conditions:
\begin{subequations}
  \begin{align}
      &  \frac{ \partial  F_{\beta}}{\partial   R }=0,  &&  \frac{ \partial  F_{\beta}}{\partial   r }=0,  &&   \frac{ \partial  F_{\beta}}{\partial   q_{1} }=0 ,   &&   \frac{ \partial  F_{\beta}}{\partial   q_{2} }=0,  \\
      &  \frac{ \partial  F_{\beta}}{\partial  T_{1}}=0,  &&   \frac{ \partial  F_{\beta}}{\partial   T_{2} }=0,  &&   \frac{ \partial  F_{\beta}}{\partial   \tau_{1} }=0  ,  &&  \frac{ \partial  F_{\beta}}{\partial   \tau_{2} }=0 .
  \end{align}
\end{subequations}
Similarly, for conjugated order parameters $\mathcal{\hat{O}}$, the following stationary conditions should be satisfied as:
 \begin{subequations}
  \begin{align}
      &   \frac{\partial F_{\beta}}{\partial \hat{R}}=0 , &&  \frac{\partial  F_{\beta}}{\partial \hat{r}}=0,   && \frac{\partial  F_{\beta}}{\partial \hat{T_{1}}}=0,   &&   \frac{\partial F_{\beta}}{\partial \hat{T_{2}}}=0, \\
      &   \frac{\partial F_{\beta}}{\partial \hat{q_{1}}}=0  , &&  \frac{\partial  F_{\beta}}{\partial \hat{q_{2}}}=0 , &&  \frac{\partial  F_{\beta}}{\partial \hat{\tau_{1}}}=0 ,  &&\frac{\partial F_{\beta}}{\partial \hat{\tau_{2}}}=0 .
    \end{align}
\end{subequations}

We first evaluate the self-consistent equations those non-conjugated order-parameters obey. For $R$, we have the following equation as
\begin{equation}
    \frac{ \partial F_{\beta} }{\partial \hat{R}}=-R+\left[\int D\mathbf{z}   \frac{\partial \ln{Z_{eff}}}{\partial R  }\right]_{ \xi^{1,true},\xi^{2,true} }=0.
\end{equation}
Thus the saddle-point equation of $R$ is given by
\begin{equation}
    R=[ \langle  \xi^{1}\xi^{2}      \rangle   ]_{ \mathbf{z},\xi^{1,true},\xi^{2,true} },  \label{R}
\end{equation}
where the thermal average  $ \langle \bullet  \rangle $ is computed under the partition function $Z_{{\rm eff}}$ (a two-spin
interaction partition function), and the outer average indicates the disorder average over Gaussian random variables $ \mathbf{z} $ and the distribution $P(\xi^{1,true},\xi^{2,true} ) $.

Similarly, for the order parameter $T_{1}$, we have the following equation as
\begin{equation}
    \frac{ \partial F_{\beta}    }{\partial  \hat{T_{1}}  }=-T_{1}+\int D \mathbf{z} \left[  \frac{1}{Z_{eff}}  \frac{\partial  Z_{eff}  }{\partial \hat{T_{1}}    }\right]_{\xi^{1,true},\xi^{2,true}}=0.
\end{equation}
Noting that $ \frac{ \partial Z_{eff}  }{ \partial \hat{T_{1}}  }     =\sum_{\xi^{1},\xi^{2}} \xi^{1,true}\xi^{1}     e^{b_{1}\xi^{1}+b_{2}\xi^{2}+b_{3}\xi^{1}\xi^{2} } $, we get the the final expression of $T_{1}$ as
\begin{equation}
    T_{1}=[\langle  \xi^{1}  \rangle \xi^{1,true}     ]_{ \mathbf{z},   \xi^{1,true},\xi^{2,true} }. \label{T1}
\end{equation}

The expressions of $T_{2},\tau_{1}$ and $\tau_{2} $ can be derived in the same way as follows:
\begin{equation}
    T_{2}=[\langle  \xi^{2}  \rangle \xi^{2,true}]_{ \mathbf{z}, \xi^{1,true},\xi^{2,true} }, \label{T2}
\end{equation}

\begin{equation}
    \tau_{1}=[\langle  \xi^{2}  \rangle \xi^{1,true}]_{ \mathbf{z},\xi^{1,true},\xi^{2,true}},\label{tau1}
\end{equation}

\begin{equation}
    \tau_{2}=[\langle  \xi^{1}  \rangle \xi^{2,true}]_{ \mathbf{z},\xi^{1,true},\xi^{2,true}}.\label{tau2}
\end{equation}

Next, we turn to the saddle-point equation of $q_{1}$, i.e.,
\begin{equation}
    \frac{\partial F_{\beta}   }{ \partial  \hat{q_{1}} }=\frac{1}{2}(q_{1}-1)+\int D\mathbf{z} \left[  \frac{1}{Z_{eff}} \frac{\partial   Z_{eff}  }{ \partial \hat{q}_{1}  } \right]_{\xi^{1,true},\xi^{2,true}}=0.
\end{equation}
Noticing that $ \frac{\partial  Z_{eff}}{\partial \hat{q}_{1}} =        \frac{1}{2}(\hat{q_{1}}-\frac{\hat{r}}{2})^{-\frac{1}{2}}\sum_{\xi^{1},\xi^{2}} \xi^{1}z_{1}e^{b_{1}\xi^{1}+b_{2}\xi^{2}+b_{3}\xi^{1}\xi^{2} } $, we get the expression of $ q_{1}$ as
\begin{equation}
    q_{1}-1+(\hat{q_{1}}-\frac{\hat{r}}{2} )^{-\frac{1}{2}} [  \langle  \xi^{1}  \rangle  z_{1}  ]_{ \mathbf{z},\xi^{1,true},\xi^{2,true}}=0.
\end{equation}

To proceed, we use the following identity 
\begin{equation}
\label{IBP}
    \int Dz f(z)z =\int Dz f^{'}(z),
\end{equation}
where $f(z)$ is any differentiable function of $z$. Thus we have the following equality as
\begin{equation}
\label{xi1z}
    [\langle   \xi^{1} \rangle z_{1}]_{\mathbf{z}}=\left[\frac{\partial}{\partial z_{1}}   \left( \frac{ \sum_{\xi^{1},\xi^{2}}     \xi^{1}   e^{b_{1}\xi^{1}+b_{2}\xi^{2}+b_{3}\xi^{1}\xi^{2}}       }{ Z_{eff}}  \right)\right]_{\mathbf{z}}=\sqrt{ \hat{q_{1}}-\frac{\hat{r}}{2}}[1- \langle \xi^{1} \rangle^{2}  ]_{\mathbf{z}}.
\end{equation}
Finally, the expression of $q_{1}$ is given by 
\begin{equation}
    q_{1}=[ \langle \xi^{1}  \rangle^{2}     ]_{ \mathbf{z},\xi^{1,true},\xi^{2,true}}. \label{q1}
\end{equation}

Similarly,  $q_{2}$  should obey the following equation given by
\begin{equation}
 q_{2}=[ \langle \xi^{2}  \rangle^{2}     ]_{\mathbf{z},\xi^{1,true},\xi^{2,true}}.   \label{q2}
\end{equation}

Following the same spirit, we get the following stationary condition for $r$ as
\begin{equation}
    \frac{r}{2}+\int D\mathbf{z} \big[ \frac{\partial}{\partial \Hat{r}} \ln{Z_{eff}}      \big]_{\xi^{1,true},\xi^{2,true}}=0.
\end{equation}
Note that
\begin{equation}
\begin{split}
   \frac{\partial}{\partial \Hat{r}} \ln{Z_{eff}}& =-\frac{1}{4}(\hat{q_{1}}-\frac{\hat{r}}{2})^{-\frac{1}{2}}\langle \xi^{1}\rangle z_{1}+\frac{1}{4}\left( \frac{\hat{r}}{2} \right )^{-\frac{1}{2}} \langle \xi^{1} \rangle  z_{3}\\
   &-  \frac{1}{4}(\hat{q_{2}}-\frac{\hat{r}}{2})^{-\frac{1}{2}}\langle \xi^{2}  \rangle z_{2}
   +   \frac{1}{4}\left( \frac{\hat{r}}{2}\right)^{-\frac{1}{2}} \langle \xi^{2}\rangle z_{3}
   - \frac{1}{2} \langle \xi^{1}\xi^{2}  \rangle. 
   \end{split}
\end{equation}
By applying Eq.~(\ref{IBP}), we can obtain the following three identities as 
\begin{equation}
    \begin{split}
        & [ \langle \xi^{2} \rangle z_{2}]_{\mathbf{z}} =\sqrt{\hat{q_{2}}-\frac{\hat{r}}{2}} \big (1-[\langle \xi^{2}       \rangle^{2} ]_{\mathbf{z}} \big),  \\
        &[\langle \xi^{1}   \rangle z_{3}]_{\mathbf{z}}=\sqrt{\frac{\hat{r}}{2}} \big( 1-[\langle \xi^{1} \rangle^{2}]_{\mathbf{z}} +[\langle \xi^{1}\xi^{2} \rangle]_{\mathbf{z}} -[\langle \xi^{1} \rangle  \langle \xi^{2}  \rangle  ]_{\mathbf{z}}   \big ), \\
        &  [\langle \xi^{2}   \rangle z_{3}]_{\mathbf{z}}=\sqrt{\frac{\hat{r}}{2}} \big (1-[\langle \xi^{2} \rangle^{2}]_{\mathbf{z}} +[\langle \xi^{1}\xi^{2} \rangle]_{\mathbf{z}} -[\langle \xi^{1} \rangle  \langle \xi^{2}  \rangle ]_{\mathbf{z}} \big).
    \end{split}
\end{equation}
Using the above three identities together with Eq.~(\ref{xi1z}), we get the expression of the saddle-point equation for $r$ as follows
\begin{equation}
    r= [\langle \xi^{1}  \rangle  \langle \xi^{2}  \rangle     ]_{\mathbf{z},\xi^{1,true},\xi^{2,true}}. \label{r}
\end{equation}

Given the result that $Z_{eff}=2e^{b_{3}}\cosh{(b_{1}+b_{2})} +2e^{-b_{3}}\cosh{(b_{1}-b_{2})}$, the thermal average like
$ \langle \xi^{1} \rangle,\langle \xi^{2} \rangle   $, and $\langle \xi^{1}\xi^{2}   \rangle $ can be easily calculated as follows:
\begin{equation}
    \begin{split}
     \langle \xi^{1} \xi^{2} \rangle_{Z_{eff}}&= \frac{\partial}{\partial b_{3}} \ln Z_{eff}    \\
     &=\frac{ e^{b_{3}}\cosh{(b_{1}+b_{2})}-e^{-b_{3}}\cosh{(b_{1}-b_{2})}}{e^{b_{3}}\cosh{(b_{1}+b_{2})}+e^{-b_{3}}\cosh{(b_{1}-b_{2}})}\\
    &=\frac{e^{b_{3}}(\cosh{b_{1}}\cosh{b_{2}}+\sinh{b_{1}}\sinh{b_{2}}) - e^{-b_{3}}(\cosh{b_{1}}\cosh{b_{2}}-\sinh{b_{1}}\sinh{b_{2}})      }{e^{b_{3}}(\cosh{b_{1}}\cosh{b_{2}}+\sinh{b_{1}}\sinh{b_{2}}) +  e^{-b_{3}}(\cosh{b_{1}}\cosh{b_{2}}-\sinh{b_{1}}\sin{b_{2}})} , \\
    &=\frac{\sinh{b_{3}}\cosh{b_{1}}\cosh{b_{2}}+\cosh{b_{3}}\sinh{b_{1}}\sinh{b_{2}} }{\cosh{b_{3}}\cosh{b_{1}}\cosh{b_{2}}+\sinh{b_{3}}\sinh{b_{1}}\sinh{b_{2}}}\\
    &=\frac{\tanh{b_{3}}+\tanh{b_{1}}\tanh{b_{2}}}{1+\tanh{b_{1}}\tanh{b_{2}}\tanh{b_{3}}},
    \end{split}  \label{xi12}
\end{equation}
and
\begin{equation}
    \begin{split}
      \langle \xi^{1} \rangle_{Z_{eff}}&= \frac{\partial}{\partial b_{1}} \ln Z_{eff}  \\
      &=\frac{ e^{b_{3}}\sinh{(b_{1}+b_{2})}+e^{-b_{3}}\sinh{(b_{1}-b_{2})}}{e^{b_{3}}\cosh{(b_{1}+b_{2})}+e^{-b_{3}}\cosh{(b_{1}-b_{2}})}\\
      &=\frac{e^{b_{3}}(\sinh{b_{1}}\cosh{b_{2}}+\cosh{b_{1}}\sinh{b_{2}}) +e^{-b_{3}}(\sinh{b_{1}}\cosh{b_{2}}-\cosh{b_{1}}\sinh{b_{2}})      }{e^{b_{3}}(\cosh{b_{1}}\cosh{b_{2}}+\sinh{b_{1}}\sinh{b_{2}}) +  e^{-b_{3}}(\cosh{b_{1}}\cosh{b_{2}}-\sinh{b_{1}}\sinh{b_{2}})} \\
     &=\frac{\cosh{b_{3}}\sinh{b_{1}}\cosh{b_{2}}+\sinh{b_{3}}\cosh{b_{1}}\sinh{b_{2}} }{\cosh{b_{3}}\cosh{b_{1}}\cosh{b_{2}}+\sinh{b_{3}}\sinh{b_{1}}\sinh{b_{2}}}\\
     &=\frac{\tanh{b_{1}}+\tanh{b_{2}}\tanh{b_{3}}}{1+\tanh{b_{1}}\tanh{b_{2}}\tanh{b_{3}}},
    \end{split}  \label{xi1} 
\end{equation}
and finally
\begin{equation}
    \begin{split}
      \langle \xi^{2}  \rangle_{Z_{eff}}&= \frac{\partial}{\partial b_{2}} \ln Z_{eff} \\
      &=\frac{ e^{b_{3}}\sinh{(b_{1}+b_{2})}-e^{-b_{3}}\sinh{(b_{1}-b_{2})}}{e^{b_{3}}\cosh{(b_{1}+b_{2})}+e^{-b_{3}}\cosh{(b_{1}-b_{2}})}\\
     &=\frac{e^{b_{3}}(\sinh{b_{1}}\cosh{b_{2}}+\cosh{b_{1}}\sinh{b_{2}}) -e^{-b_{3}}(\sinh{b_{1}}\cosh{b_{2}}-\cosh{b_{1}}\sinh{b_{2}})      }{e^{b_{3}}(\cosh{b_{1}}\cosh{b_{2}}+\sinh{b_{1}}\sinh{b_{2}}) +  e^{-b_{3}}(\cosh{b_{1}}\cosh{b_{2}}-\sinh{b_{1}}\sinh{b_{2}})} \\
     &=\frac{\cosh{b_{2}}\sinh{b_{1}}\sinh{b_{3}}+\sinh{b_{2}}\cosh{b_{1}}\cosh{b_{3}} }{\cosh{b_{3}}\cosh{b_{1}}\cosh{b_{2}}+\sinh{b_{3}}\sinh{b_{1}}\sinh{b_{2}}}\\
     &=\frac{\tanh{b_{2}}+\tanh{b_{1}}\tanh{b_{3}}}{1+\tanh{b_{1}}\tanh{b_{2}}\tanh{b_{3}}}.
    \end{split}  \label{xi2}
\end{equation}

In case of $\hat{r}<0$, we can re-parameterize $b_{1}$ and $b_{2}$ as
 \begin{subequations}
   \begin{align}
       &   b_{1}=\sqrt{\hat{q_{1}}}z_{1}+\hat{T_{1}} \xi^{1,true}  +\hat{\tau_{2}}\xi^{2,true},\\
       & b_{2}=\sqrt{\hat{ q_{2}}} \big( \psi z_{1}+\sqrt{1-\psi^{2}}z_{2} \big )+\hat{T_{2}}\xi^{2,true}+\hat{\tau_{1}}\xi^{1,true}, \\
       &  \mathcal{\psi}= \frac{\hat{r}}{2\sqrt{\hat{q_{1}} \hat{q_{2}}} }.
   \end{align}
 \end{subequations}
We remark that this re-parameterization does not change the final
results of multidimensional Gaussian integrations in the saddle-point equation.

To sum up, the saddle-point equations for non-conjugated order parameters are given by
\begin{subequations}
\label{nconjop}
  \begin{align}
      T_{1}&=[\xi^{1,true}\langle \xi^{1} \rangle ]_{\mathbf{z},\xi^{1,true},\xi^{2,true}} , \\
      T_{2}&=[\xi^{2,true}\langle  \xi^{2} \rangle ]_{\mathbf{z},\xi^{1,true},\xi^{2,true}},\\
      q_{1}&=[  \langle  \xi^{1}  \rangle^{2}  ]_{\mathbf{z},\xi^{1,true},\xi^{2,true}}, \\
      q_{2}&=[  \langle  \xi^{2}  \rangle^{2}  ]_{\mathbf{z},\xi^{1,true},\xi^{2,true}},  \\
      \tau_{1}&=[ \xi^{1,true} \langle  \xi^{2} \rangle ]_{\mathbf{z},\xi^{1,true},\xi^{2,true}} , \\
      \tau_{2}&=[ \xi^{2,true} \langle  \xi^{1} \rangle ]_{\mathbf{z},\xi^{1,true},\xi^{2,true} },  \\
      R&=[\langle \xi^{1}\xi^{2} \rangle]_{\mathbf{z}, \xi^{1,true},\xi^{2,true}}, \\
      r&=[\langle \xi^{1}\rangle \langle \xi^{2} \rangle ]_{\mathbf{z}, \xi^{1,true},\xi^{2,true}}.
  \end{align}
\end{subequations}

Next, we derive the saddle-point equations for those conjugated order parameters.
For $\Hat{R}$, we obtain the saddle point equation as
\begin{equation}
    \frac{\partial   F_{\beta}   }{\partial R    }=-\Hat{R}-\alpha\beta^{2}\tanh{(\beta^{2}R)}+\frac{\alpha e^{-\beta^{2}} }{\cosh{(\beta^{2}q)}   }\int  D\mathbf{t}    \cosh{\beta t_{0}}\cosh\beta(qt_{0}+\sqrt{1-q^{2}}x_{0} ) \frac{\partial   }{\partial R  }\ln{  Z_{E}}=0,
\end{equation}
where $\frac{\partial   }{\partial R  }\ln{  Z_{E}}=\beta^{2}\frac{ e^{\beta^{2}(R-r)} \cosh{(\beta \Lambda_{+})}- e^{-\beta^{2}(R-r)} \cosh{(\beta \Lambda_{-}}  ) }{ e^{\beta^{2}(R-r)} \cosh{(\beta \Lambda_{+})}+ e^{-\beta^{2}(R-r)} \cosh{(\beta \Lambda_{-})}   }=\beta^{2}G_{c}^{-}.$
Therefore, the saddle-point equation of $ \Hat{R} $ is given by
 \begin{equation}
    \Hat{R}=\frac{\alpha  \beta^{2} e^{-\beta^{2}} }{\cosh{(\beta^{2}q})   }\int D\mathbf{t} [\cosh{\beta t_{0}}\cosh\beta(qt_{0}+\sqrt{1-q^{2}}x_{0} ) G_{c}^{-}  -\alpha\beta^{2}\tanh{(\beta^{2}R)} .
\end{equation}
For convenience, we define the measure  $\langle  \bullet \rangle $ as $ \frac{e^{-\beta^{2}}}{\cosh{(\beta^{2}}q)} \int D\mathbf{t}    \cosh{\beta t_{0}}\cosh{\beta (qt_{0}+\sqrt{1-q^{2}}x_{0})}  \bullet $. 
As a result,
\begin{equation}
    \Hat{R}=\alpha \beta^{2}\langle G_{c}^{-} \rangle -\alpha \beta^{2}\tanh{(\beta^{2}R)}. 
\end{equation}

For $ \Hat{T_{1}}$, we have the following condition 
  \begin{equation}
  \label{eqT1hat}
      \frac{\partial  F_{\beta}   }{\partial  T_{1}   }=-\Hat{T}_{1}+\frac{\alpha e^{-\beta^{2}}  }{\cosh{(\beta^{2}q})}\int D\mathbf{t} \cosh{\beta t_{0}}\cosh{\beta(qt_{0}+\sqrt{1-q^{2}}x_{0})} \frac{\partial }{\partial T_{1}}\ln{Z_{E} }=0.
  \end{equation}
To proceed, we first get the derivation of $\Lambda_{+}$ and $\Lambda_{-}$  w.r.t $T_{1}$ as follows
 \begin{subequations}
   \begin{align}
       \frac{\partial  \Lambda_{+}  }{\partial T_{1}  }&=t_{0}-\frac{q}{\sqrt{1-q^{2}}}x_{0}+\frac{\partial}{\partial T_{1}}\left( B+\frac{r-A}{B}\right)u+ \frac{\partial K}{\partial T_{1}}u^{'},\\
       \frac{\partial  \Lambda_{-}  }{\partial T_{1}  }&=t_{0}-\frac{q}{\sqrt{1-q^{2}}}x_{0}+\frac{\partial}{\partial T_{1}}\left( B-\frac{r-A}{B}\right)u- \frac{\partial K}{\partial T_{1}}u^{'}.
   \end{align}
 \end{subequations}
Then, the derivation of $\ln{Z_{E}} $ w.r.t $T_{1}$ can be simplified into the form as
 \begin{equation}
     \frac{\partial \ln{ Z_{E}} }{ \partial T_{1}}=\beta\left[ G_{s}^{+}t_{0}-\frac{q}{\sqrt{1-q^{2}}}G_{s}^{+}x_{0}+ \frac{\partial  B}{\partial T_{1}}G_{s}^{+}u+  \frac{ \partial  }{\partial T_{1} }\left(\frac{r-A}{B}\right)G_{s}^{-}u+\frac{\partial K }{\partial T_{1}} G^{-}_{s} u^{'}\right].
 \end{equation}
To further simplify the result, we need to evaluate the following equations. The first one is derived by applying Eq.~(\ref{IBP}) as
\begin{equation}
\label{gpt0}
    \begin{split}
        &\int D\mathbf{t}  \cosh{\beta t_{0}}\cosh{\beta(qt_{0}+\sqrt{1-q^{2}}x_{0})}G_{s}^{+}t_{0}\\
        &=\int D\mathbf{t}   \frac{\partial}{\partial t_{0}} \bigg(   \cosh{\beta t_{0}}\cosh{\beta(qt_{0}+\sqrt{1-q^{2}}x_{0})}G_{s}^{+}\bigg )\\
        &=\beta \int D\mathbf{t} \bigg[  \sinh{\beta t_{0}}\cosh{\beta(qt_{0}+\sqrt{1-q^{2}}x_{0})}+q\cosh{\beta t_{0}}\sinh{\beta(qt_{0}+\sqrt{1-q^{2}}x_{0})} \bigg]  G_{s}^{+} \\
        &+\beta   \int D\mathbf{t}  \cosh{\beta t_{0}}\cosh{\beta(qt_{0}+\sqrt{1-q^{2}}x_{0})} \bigg[ T_{1}+\tau_{1}G_{c}^{-}-T_{1}(G_{s}^{+})^{2}-\tau_{1}G_{s}^{+}G_{s}^{-}  \bigg].
    \end{split}
\end{equation}
The second one is derived as
\begin{equation}
\label{gpx0}
    \begin{split}
        &\int D\mathbf{t} \cosh{\beta t_{0}}\cosh{\beta(qt_{0}+\sqrt{1-q^{2}}x_{0})}G_{s}^{+}x_{0}\\
        &=\int D\mathbf{t}   \frac{\partial}{\partial x_{0}} \bigg( \cosh{\beta t_{0}}\cosh{\beta(qt_{0}+\sqrt{1-q^{2}}x_{0})}G_{s}^{+} \bigg )\\
       &=\beta\sqrt{1-q^{2}}\int D\mathbf{t}  \cosh{\beta t_{0}}\sinh{\beta(qt_{0}+\sqrt{1-q^{2}}x_{0})}G_{s}^{+}\\
       &+ \frac{\beta}{\sqrt{1-q^{2}}}\int D\mathbf{t}     
           \cosh{\beta t_{0}}\cosh{\beta(qt_{0}+\sqrt{1-q^{2}}x_{0})} \\
       &\times\bigg[(\tau_{2}-qT_{1})+(T_{2}-q\tau_{1})G_{c}^{-}-(\tau_{2}-qT_{1})(G_{s}^{+})^{2}-(T_{2}-q\tau_{1})G_{s}^{+}G_{s}^{-} \bigg].
    \end{split}
\end{equation}
The third one is derived as
\begin{equation}
\label{gpu}
    \begin{split}
        &\int D\mathbf{t}  \cosh{\beta t_{0}}\cosh{\beta(qt_{0}+\sqrt{1-q^{2}}x_{0})}G_{s}^{+}u\\
        &=\int D\mathbf{t}   \frac{\partial}{\partial u} \bigg(   \cosh{\beta t_{0}}\cosh{\beta(qt_{0}+\sqrt{1-q^{2}}x_{0})}G_{s}^{+}   \bigg )\\
        &=\beta \int D\mathbf{t} \cosh{\beta t_{0}}\cosh{\beta(qt_{0}+\sqrt{1-q^{2}}x_{0})}\bigg[  B+\frac{r-A}{B}G_{c}^{-}-B(G_{s}^{+})^{2}-\frac{r-A}{B}G_{s}^{+}G_{s}^{-}  \bigg].
    \end{split}
\end{equation}
The fourth one is derived as
\begin{equation}
\label{gmu}
    \begin{split}
        &\int D\mathbf{t}  \cosh{\beta t_{0}}\cosh{\beta(qt_{0}+\sqrt{1-q^{2}}x_{0})}G_{s}^{-}u\\
        &=\int D\mathbf{t}   \frac{\partial}{\partial u} \bigg(   \cosh{\beta t_{0}}\cosh{\beta(qt_{0}+\sqrt{1-q^{2}}x_{0})}G_{s}^{-}   \bigg )\\
        &=\beta \int D\mathbf{t} \cosh{\beta t_{0}}\cosh{\beta(qt_{0}+\sqrt{1-q^{2}}x_{0})}\bigg[  BG_c^-+\frac{r-A}{B}-\frac{r-A}{B}(G_{s}^{-})^{2}-BG_{s}^{+}G_{s}^{-}  \bigg].
    \end{split}
\end{equation}
The last one is given by
\begin{equation}
\label{gmup}
    \begin{split}
        & \int D\mathbf{t}  \cosh{\beta t_{0}}\cosh{\beta(qt_{0}+\sqrt{1-q^{2}}x_{0})}G_{s}^{-}u^{'} \\
        &=\int D\mathbf{t}   \frac{\partial}{\partial u^{'}} \bigg(   \cosh{\beta t_{0}}\cosh{\beta(qt_{0}+\sqrt{1-q^{2}}x_{0})}G_{s}^{-}  \bigg )\\
        &=\beta  K \int D\mathbf{t}  \cosh{\beta t_{0}}\cosh{\beta(qt_{0}+\sqrt{1-q^{2}}x_{0})}\bigg[ 1-(G_{s}^{-})^{2} \bigg].
    \end{split}
\end{equation}
Through a bit lengthy algebraic manipulations, we get from Eq.~(\ref{eqT1hat})
 \begin{equation}
     \Hat{T}_{1}=\frac{\alpha  \beta^{2}e^{-\beta^{2}} }{\cosh{(\beta^{2}q})}\int D\mathbf{t} \sinh{\beta t_{0}}\cosh{\beta(qt_{0}+\sqrt{1-q^{2}}x_{0})}G_{s}^{+}.
     \end{equation} 
We thus define another measure $\langle \langle   \bullet  \rangle \rangle =  \frac{e^{-\beta^{2}}}{\cosh{(\beta^{2}q})}\int D\mathbf{t} \sinh{\beta t_{0}} \cosh{\beta(qt_{0}+\sqrt{1-q^{2}}x_{0} )}  \bullet  $, 
and it then follows that
\begin{equation}
    \Hat{T_{1}}=\alpha \beta^{2}\langle  \langle  G_{s}^{+} \rangle\rangle.\label{hatT1}
\end{equation}

Similarly, we can obtain the saddle-point equation of $\hat{\tau_{1}}$ as
\begin{equation}
    \hat{\tau}_{1}=\alpha \beta^{2}\langle\langle G_{s}^{-}   \rangle\rangle.  \label{hattau1}
\end{equation}

Next we turn to the saddle-point equations for $\hat{T}_{2}$ and $ \Hat{\tau_{2}}$. We first get the derivation of $\Lambda_{+}$ and $\Lambda_{-} $ w.r.t $T_2$ as
\begin{subequations}
  \begin{align}
      &   \frac{\partial \Lambda_{+} }{\partial T_{2} }=\frac{ x_{0}}{\sqrt{1-q^{2}}}-\frac{1}{B} \frac{\partial A}{\partial T_{2}}u+\frac{\partial K   }{\partial T_{2}}u^{'} , \\
      &  \frac{\partial \Lambda_{-} }{\partial T_{2} }=-\frac{ x_{0}}{\sqrt{1-q^{2}}}+\frac{1}{B} \frac{\partial A}{\partial T_{2}}u-\frac{\partial K   }{\partial T_{2}}u^{'}.
  \end{align}
\end{subequations}
Based on the above equations, we get the derivation of $\ln{Z_{E}} $ w.r.t $ T_{2} $ given by
\begin{equation}
    \frac{\partial \ln{ Z_{E} } }{\partial T_{2} }=\beta\left[ \frac{x_{0}}{\sqrt{1-q^{2}}}G_{s}^{-}-\frac{1}{B}\frac{\partial A  }{\partial T_{2}  }G_{s}^{-}u+\frac{\partial K  }{\partial T_{2}  }G_{s}^{-}u^{'} \right].
\end{equation}
Then we have 
\begin{equation}
    \Hat{T}_{2}=\frac{\alpha \beta  e^{-\beta^{2}}  }{\cosh{(\beta^{2}q)} }\int D\mathbf{t}\cosh{\beta t_{0}}\cosh{\beta(qt_{0}+\sqrt{1-q^{2}}x_{0})} \bigg[ \frac{x_{0}}{\sqrt{1-q^{2}}}G_{s}^{-}-\frac{1}{B} \frac{ \partial A  }{  \partial T_{2}} G_{s}^{-}u+\frac{\partial K  }{\partial T_{2}} G_{s}^{-}u' \bigg] .
\end{equation}
For a further simplification, we need to derive the following identity as
\begin{equation}
\label{gmx0}
    \begin{split}
      & \int D\mathbf{t}  \cosh{\beta t_{0}}\cosh{\beta(qt_{0}+\sqrt{1-q^{2}}x_{0})}G_{s}^{-}x_{0} \\
      &= \int D\mathbf{t}   \frac{\partial}{\partial x_{0}} \bigg(\cosh{\beta t_{0}}\cosh{\beta(qt_{0}+\sqrt{1-q^{2}}x_{0})}G_{s}^{-} \bigg )\\
      &=\beta\sqrt{1-q^{2}}\int D\mathbf{t}  \cosh{\beta t_{0}}\sinh{\beta(qt_{0}+\sqrt{1-q^{2}}x_{0})}G_{s}^{-}\\
      &+ \frac{\beta}{\sqrt{1-q^{2}}}\int D\mathbf{t}  
     \cosh{\beta t_{0}}\cosh{\beta(qt_{0}+\sqrt{1-q^{2}}x_{0})}\\
      &\times\bigg[ (  \tau_{2}-qT_{1})G_{c}^{-}-(\tau_{2}-qT_{1})G_{s}^{+}G_{s}^{-}-(T_{2}-q\tau_{1})(G_{s}^{-})^{2}+(T_{2}-q\tau_{1}) \bigg ].
    \end{split}
\end{equation}
Using Eq.~(\ref{gmx0}) together with Eq.~(\ref{gmu}) and Eq.~(\ref{gmup}), we finally arrive at 
the saddle-point equation of $\hat{T}_{2}  $:
\begin{equation}
    \Hat{T}_{2}=\frac{\alpha \beta^{2}e^{-\beta^{2}}     }{ \cosh{(\beta^{2}q})  }\int D\mathbf{t}  \cosh{\beta t_{0}}\sinh{\beta(qt_{0}+\sqrt{1-q^{2}}x_{0})}   G_{s}^{-}.  
\end{equation}
We thus define the third measure $\langle\langle\langle \bullet \rangle\rangle\rangle=\frac{ e^{-\beta^{2}}}{\cosh{(\beta^{2}q})}\int D\mathbf{t}   \cosh{\beta t_{0}}\sinh{\beta(qt_{0}+\sqrt{1-q^{2}}x_{0})}  \bullet  $.
We then write the saddle-point equation in a compact form as
\begin{equation}
    \hat{T_{2}}=\alpha\beta^{2}\langle\langle\langle  G_{s}^{-}  \rangle\rangle\rangle.  \label{hatT2}
\end{equation}

Similarly, we obtain the saddle-point equation for $\hat{\tau_{2}}  $ as
\begin{equation}
    \hat{\tau_{2}}=\alpha\beta^{2}\langle\langle\langle  G_{s}^{+}  \rangle\rangle\rangle.  \label{hattau2}
\end{equation}

Then we turn to the saddle-point equations of $\Hat{q}_{1}$ and $\Hat{q}_{2}$. From $\frac{\partial F_{\beta}  }{\partial q_{1}   }=0  $, we get
\begin{equation}
    \frac{1}{2}\Hat{q}_{1}-\frac{\alpha \beta^{2}}{2}+\frac{\alpha \beta e^{-\beta^{2}}  }{ \cosh{(\beta^{2}q}) }\int D\mathbf{t}  \cosh{\beta t_{0}}\cosh{\beta(qt_{0}+ \sqrt{1-q^{2}}x_{0})}  \frac{\partial  \ln{Z_{E}}    }{\partial q_{1}}=0.
\end{equation}
The derivation of $\ln{Z_{E}}$ w.r.t $q_{1}$ is given by
\begin{equation}
    \frac{ \partial  \ln{Z_{E}}    }{\partial q_{1}  }=\frac{\partial B  }{ \partial  q_{1}   }G_{s}^{+}u+\frac{\partial}{\partial q_{1}}\left(\frac{r-A}{B}\right)G_{s}^{-}u+\frac{   \partial K}{\partial q_{1}}G_{s}^{-}u^{'} .
    \end{equation}
Using Eq.~(\ref{gpu}), Eq.~(\ref{gmu}) and Eq.~(\ref{gmup}), we get the saddle-point equation of $ \hat{q}_{1} $ as 
\begin{equation}
    \begin{split}
        \hat{q}_{1}&= \frac{\alpha  \beta^{2} e^{-\beta^{2}} }{\cosh{(\beta^{2}q})}\int D\mathbf{t} \cosh{(\beta t_{0})}\cosh{\beta( qt_{0}+\sqrt{1-q^{2}}x_{0})}( G_{s}^{+} )^{2}\\
        &=\alpha \beta^{2}\langle  (G_{s}^{+})^{2} \rangle. \label{hatq1}
    \end{split}
\end{equation}

Similarly, we can derive the saddle-point equation for $\hat{q}_2$ as
\begin{equation}
    \begin{split}
        \hat{q}_{2}&= \frac{\alpha  \beta^{2} e^{-\beta^{2}} }{\cosh{(\beta^{2}q})}\int D\mathbf{t} \cosh{(\beta t_{0})}\cosh{\beta( qt_{0}+\sqrt{1-q^{2}}x_{0})}( G_{s}^{-} )^{2}\\
        &=\alpha \beta^{2}\langle  (G_{s}^{-})^{2} \rangle. \label{hatq2}
    \end{split}
\end{equation}

Lastly, we derive the saddle-point equation for $\hat{r}$ as
\begin{equation}
    \frac{\hat{r}}{2}+\frac{ \alpha e^{-\beta^{2}}  }{ \cosh{\beta^{2}q} }\int D\mathbf{t} \cosh{\beta t_{0}}\cosh{\beta(qt_{0}+\sqrt{1-q^{2}}x_{0} )}  \frac{\partial  \ln{Z_{E}}    }{\partial r }=0.
\end{equation}
Noting that $ \frac{\partial  \ln{Z_{E}}}{\partial  r}=-\beta^{2}G_{c}^{-}+\beta\left( \frac{1}{B}G_{s}^{-}u+\frac{\partial K}{\partial r}G_{s}^{-}u^{'} \right)$, we get the saddle-point equation of $\Hat{r}$ as
\begin{equation}
    \Hat{r}=2\alpha \beta^{2}\langle G_{s}^{+}G^{-}_{s}  \rangle. \label{hatr}
\end{equation}
Note that Eq.~(\ref{gmu}) and Eq.~(\ref{gmup}) are used to derive the final result.

To sum up, the saddle-point equations of our minimal model are listed as follows
\begin{subequations}
\label{conjop}
  \begin{align}
      &  \hat{T_{1}}= \alpha \beta^{2}   \langle  \langle  G_{s}^{+}  \rangle \rangle, \\
      &  \hat{T_{2}}= \alpha \beta^{2}    \langle \langle \langle G_{s}^{-} \rangle\rangle \rangle,  \\
      &  \hat{\tau_{1}}=\alpha  \beta^{2}    \langle \langle    G_{s}^{-}  \rangle\rangle, \\
      &   \hat{\tau_{2}}=\alpha \beta^{2}\langle  \langle  \langle  G_{s}^{+} \rangle  \rangle \rangle,\\
      &  \hat{q_{1}}=\alpha  \beta^{2}  \langle  (G_{s}^{+})^{2}  \rangle,\\
      &\hat{q_{2}}=  \alpha \beta^{2} \langle   (G_{s}^{-})^{2}  \rangle , \\
      &   \hat{r}=2\alpha \beta^{2} \langle   G_{s}^{+}G_{s}^{-}  \rangle , \\
      &  \Hat{R}=  \alpha \beta^{2}  \langle  G_{c}^{-}   \rangle- \alpha \beta^{2}\tanh{(\beta^{2}R)}  .
  \end{align}
\end{subequations}
\section{The free energy function in the limit of $q=0$}
\label{app4}
In the case of $q=0$, the saddle point equation of the minimal model has the solution: $q_1=q_2=T_1=T_2$ and other order parameters vanish.
Thus, we can simplify $\Lambda_+$ and $\Lambda_{-}$ as follows,
\begin{subequations}\label{app4-1}
  \begin{align}
      & \Lambda_{+}=T_{1}t_{0}+T_{2}x_{0}+\sqrt{q_{1}-(T_{1})^{2}}u+\sqrt{q_{2}-(T_{2})^{2}}u', \\
      &  \Lambda_{-}=T_{1}t_{0}-T_{2}x_{0}+\sqrt{q_{1}-(T_{1})^{2}}u-\sqrt{q_{2}-(T_{2})^{2}}u'.
  \end{align}
\end{subequations}
We then define $ \chi_{1}=T_{1}t_{0}+\sqrt{q_{1}- (T_{1})^{2}}u $, and  $\chi_{2}=T_{2}x_{0}+\sqrt{q_{2}- (T_{2})^{2}}u' $, so the saddle point eqtation of $ \hat{T}_{1} $ is given by
\begin{equation}\label{app4-2}
    \begin{split}
         \hat{T}_{1}&=\alpha  \beta^{2}e^{-\beta^{2}}\int D\mathbf{t}  \sinh{\beta  t_{0}} \cosh{\beta x_{0}} \bigg [ \frac{ \sinh{\beta \Lambda_{+}}+\sinh{\beta \Lambda_{-} } }{\cosh{\beta \Lambda_{+}}+\cosh{\beta \Lambda_{-}}}  \bigg]   \\
        &= \alpha  \beta^{2}e^{-\beta^{2}}\int D\mathbf{t}  \sinh{\beta  t_{0}} \cosh{\beta x_{0}}     \bigg [\frac{\sinh{\beta \chi_{1}} \cosh{\beta \chi_{2}}   }{ \cosh{\beta  \chi_{1}} \cosh{\beta  \chi_{2}}  }  \bigg]  \\
        &=\alpha  \beta^{2}e^{- \frac{\beta^{2}}{2} }\int Dt_{0}Du  \sinh{\beta t_{0}}\tanh{\beta(T_{1}t_{0}+\sqrt{q_{1}-(T_{1})^{2}}        u  )},
    \end{split}
\end{equation}
where we used the identity $\int Dx_0\cosh(\beta x_0)=e^{\beta^2/2}$. Similarly, one can prove that $\hat{T}_1=\hat{T}_2$. As for $\Hat{q}_{1}$,  we will have 
\begin{equation}\label{app4-3}
    \begin{split}
        \Hat{q}_{1}&=\alpha \beta^{2}e^{-\beta^{2}}\int D\mathbf{t} \cosh{\beta t_{0}}\cosh{\beta x_0} \bigg[  \frac{\sinh{\beta \Lambda_{+}}+\sinh{\beta \Lambda_{-}}}{\cosh{\beta \lambda_{+}}+\cosh{\beta \Lambda_{-}}  }  \bigg]  ^{2}  \\
        &=\alpha \beta^{2}e^{-\beta^{2}}\int D\mathbf{t} \cosh{\beta t_{0}}\cosh{\beta x_{0}} \bigg[ \frac{ \sinh{\beta\chi_{1}}  \cosh{\beta \chi_{2}}      }{ \cosh{\beta  \chi_{1}}\cosh{\beta \chi_{2}}             }  \bigg]^{2}\\
        &=\alpha \beta^{2} e^{ -\frac{\beta^{2}}{2}  }\int D t_{0}Du \cosh{\beta t_{0}}   \tanh^2{\beta( T_{1}t_{0}+\sqrt{q_{1}-(T_{1})^2}u                            )}.
        \end{split}
\end{equation}
Similarly, one can prove that $\hat{q}_1=\hat{q}_2$.

It is easy to prove that $ \Hat{\tau}_{1}=0,\Hat{\tau_{2}}=0    $ and $ \Hat{R}=0 ,\Hat{r}=0 $,  then we can  express $b_{1},b_{2} $ and $b_3$  as :
\begin{subequations}
  \begin{align}
      & b_{1}=\hat{T}_{1}\xi^{1,true}+\sqrt{\Hat{q_{1}}}z_{1}  ,\\
      &  b_{2}=\hat{T}_{2}\xi^{2,true}+\sqrt{\Hat{q}_{2}}z_{2},\\
      &  b_{3}=0.
  \end{align}
\end{subequations}
Therefore, $ \int D\mathbf{z}[\ln Z_{eff}]_{\xi^{1,true},\xi^{2,true}} $ can be simplified as $2\int Dz\ln 2\cosh(\hat{T}_1+\sqrt{\hat{q}_1}z)$.
$T_1$ becomes
\begin{equation}
    \begin{split}
        T_{1}&= \Bigg[  \int Dz_{1}Dz_{2}Dz_{3}    \xi^{1,true}  \tanh{(\Hat{T}_{1}\xi^{1,true}+\sqrt{q_{1}}z_{1} )}                          \bigg ]_{\xi^{1,true},\xi^{2,true}}  \\
        &= \int Dz_{1}   \frac{1}{2} \bigg[ \tanh{(\Hat{T}_{1}+\sqrt{\Hat{q}_{1}}z_{1}) }-\tanh{(-\Hat{T}_{1}+\sqrt{q}_{1}z_{1})}  \bigg ]     \\
        &=   \int Dz_{1}     \frac{1}{2} \bigg[     \tanh(\Hat{T}_{1}+\sqrt{\Hat{q}_{1}}z_{1} )-\tanh{(-\Hat{T}_{1}-\sqrt{q}_{1}z_{1}    )}    \bigg ]\\
        &=\int Dz_{1} \tanh{(\Hat{T}_{1}+\sqrt{q_{1}}z_{1}) }.
    \end{split}
\end{equation}
One can easily prove that $T_1=T_2$. Similarly for the order parameter $ q_{2} $, we can also get:
\begin{equation}
    \begin{split}
        q_{2}=&\bigg [ \int Dz_{2}   \tanh^2{(\Hat{T}_{2}\xi^{1,true}+\sqrt{ \Hat{q}_{2} }z_{1}   )}  \bigg ]_{\xi^{1,true},\xi^{2,true} }  \\
        &=\frac{1}{2} \int Dz_{2} \bigg[ \tanh^2{(\Hat{T}_{2}+ \sqrt{\Hat{q}_{2}}z_{2} )}  +  \tanh^2{(-\Hat{T}_{2}+ \sqrt{\Hat{q}_{2}}z_{2} )}    \bigg]   \\
        &= \int Dz_{2}  \tanh^2{( \Hat{T}_{2}+\sqrt{\Hat{q}_{2}}z_{2}              )}.
    \end{split}
\end{equation}
It is easy to show that $q_1=q_2$, and moreover $R=r=\tau_1=\tau_2=0$. To sum up, we recover the saddle point equations of one-bit RBM reported in Ref.~\cite{Huang-2017}.

Next, we show the $q=0$ version of the free energy function. It is easy to show that 
$Z_{{\rm E}}=\cosh{\beta(\chi_{1}+\chi_{2})}+\cosh{\beta(\chi_{1}-\chi_{2})} =2\cosh{\beta \chi_{1}}  \cosh{\beta \chi_{2}}  $. Therefore, we have the following integral
\begin{equation}
    \begin{split}
       &\alpha e^{-\beta^{2}} \int D\mathbf{t} \cosh{\beta t_{0}}\cosh{\beta x_{0}} \ln{Z_{{\rm E}}} =\alpha e^{-\beta^{2}} \int D\mathbf{t} \cosh{\beta t_{0}}\cosh{\beta x_{0}} \ln({2\cosh{\beta \chi_{1}}\cosh{\beta \chi_{2}}  })   \\
       &=\alpha \ln{2}+ 2\alpha e^{-\frac{\beta^{2}}{2}} \int Du D t_{0} \cosh{\beta t_{0}} \ln{\cosh{\beta(T_{1}t_{0}+\sqrt{q_{1}-(T_{1})^{2}}u)}}.
    \end{split}
\end{equation}
Collecting all the relevant terms, we can show that the free
energy of our minimal model with $q=0$ is two times as large as that of one-bit RBM, which can also be intuitively understood by the argument
that the partition function factorizes as $\Omega=\Omega_{{\rm one-bit-RBM}}^2$. Therefore we can conclude that the critical data size for spontaneous symmetry breaking does not change even if
an additional hidden node is added. This conclusion seems to carry over to the case of more hidden nodes following the principle of the partition function's factorization.

\section{Derivation of the critical data size (Eq.~(\ref{thre}) in the main text)}
\label{app3}
 We assume that near to the transition point, all order parameters are very small such that we can expand them to leading order.
 According to Eq.~(\ref{nconjop}), when the critical point is approached from below,
 $\langle  \xi^{1} \rangle \simeq \tanh{b_{1}}\simeq  b_{1} $. Analogously, $\langle  \xi^{2}\rangle \simeq   b_{2} $, and
 $\langle  \xi^{1}\xi^{2}  \rangle \simeq b_{3} $. We thus have the following equalities in this limit:
 \begin{eqnarray}
T_{1}&= [\xi^{1,true} \langle  \xi^{1}  \rangle   ]_{\mathbf{z},\xi^{1,true},\xi^{2,true}}=\Hat{T_{1}}+q\Hat{\tau_{2}},\\
  \tau_{2}&= [\xi^{2,true} \langle  \xi^{1}  \rangle]_{\mathbf{z},\xi^{1,true},\xi^{2,true}}=\Hat{\tau_{2}}+q \hat{T_{1}}.  
\end{eqnarray}

Similarly, in the limit of vanishing order parameters, we have the following approximation
\begin{equation}
   \begin{split}
        G_{s}^{+}&=\frac{e^{\beta^{2}(R-r)}\sinh{(\beta \Lambda_{+})}+e^{-\beta^{2}(R-r)}\sinh{(\beta \Lambda_{-})} }{ e^{\beta^{2}(R-r)}\cosh{(\beta \Lambda_{+})}+e^{-\beta^{2}(R-r)}\cosh{(\beta \Lambda_{-})}} \\
        &=\frac{\beta}{2}( \Lambda_{+}+\Lambda_{-} ).
   \end{split}
\end{equation}
Inserting this approximation into the saddle-point equations of $\hat{T}_1$ and $\hat{\tau}_2$, we
obtain the approximate results of $\Hat{T_{1}}$ and $\Hat{\tau_{2}} $ as
\begin{equation}
    \begin{split}
        \hat{T_{1}}&=\alpha\beta^2\langle\langle G_s^+\rangle\rangle\simeq\frac{\alpha \beta^{2} e^{-\beta^{2}}}{\cosh{(\beta^{2}q})}\int D\mathbf{t}\sinh{\beta t_{0}}\cosh{\beta(qt_{0}+\sqrt{1-q^{2}}x_{0})} \frac{\beta}{2}[ \Lambda_{+}+\Lambda_{-} ]\\ 
        &=\alpha \beta^{4}[T_{1}+\tanh(\beta^{2}q)\tau_{2}], \\ 
        \hat{\tau_{2}}&=\alpha\beta^2\langle\langle\langle G_s^+\rangle\rangle\rangle\simeq\frac{\alpha \beta^{2} e^{-\beta^{2}}}{\cosh{(\beta^{2}q})}\int D\mathbf{t}\cosh{\beta t_{0}}\sinh{\beta(qt_{0}+\sqrt{1-q^{2}}x_{0})} \frac{\beta}{2}[ \Lambda_{+}+\Lambda_{-} ]\\
        &=\alpha \beta^{4}[\tau_{2}+\tanh(\beta^{2}q)T_{1}].\label{hatop}
    \end{split}
\end{equation}
We recast the equations for all these four order parameters in a matrix form as
\begin{eqnarray}
\label{mat1}
  \left(
\begin{array}
{c}
T_{1}\\
\tau_{2} 
\end{array}
\right)=\left(
\begin{array}
{cc}
1 & q\\
q & 1
\end{array}
\right)\left(
\begin{array}
{c}
\Hat{T}_{1}\\
\hat{\tau_{2}}
\end{array}
\right),\\  
  \left(
\begin{array}
{c}
\Hat{T}_{1}\\
\Hat{\tau}_{2} 
\end{array}
\right)= \alpha \beta^{4}  \left(
\begin{array}
{cc}
1&\tanh{(\beta^{2}q)}\\
\tanh{(\beta^{2}q)} & 1
\end{array}
\right)\left(
\begin{array}
{c}
T_{1}\\
\tau_{2}
\end{array}
\right). \label{mat2}
\end{eqnarray}

From the Eq.~(\ref{mat1})  and Eq.~(\ref{mat2}), $T_{1}$ and $\tau_{2} $ can be calculated out as
\begin{equation}
  \left(
\begin{array}
{c}
T_{1}\\
\tau_{2}
\end{array}
\right)= \alpha \beta^{4}  \left(
\begin{array}
{cc}
  1+q\tanh{(\beta^{2}q)}     & q+\tanh{(\beta^{2}q)}        \\
  q+\tanh{(\beta^{2}q)}   &   1+q\tanh{(\beta^{2}q)}          \\

\end{array}
\right)\left(
\begin{array}
{c}
T_{1}\\
\tau_{2}\\
\end{array}
\right)=\mathcal{M}\left(
\begin{array}
{c}
T_{1}\\
\tau_{2}\\
\end{array}
\right),
\label{matrix3}
\end{equation}
where the matrix $\mathcal{M}$ is named the stability matrix,
whose largest eigenvalue determines the critical value of the learning data size $\alpha_{c}$.
In detail, the stability matrix has two eigenvalues:
\begin{eqnarray}
 \lambda_+&=\alpha\beta^4\left(1+q\tanh{(\beta^{2}q)}+|q+\tanh(\beta^{2}q)|\right),\\
 \lambda_-&=\alpha\beta^4\left(1+q\tanh{(\beta^{2}q)}-|q+\tanh(\beta^{2}q)|\right).
\end{eqnarray}
The $\alpha_c$ can be read off from $\lambda_+=1$, i.e.,
\begin{equation}
    \alpha_{c}=\frac{\beta^{-4}}{1+q\tanh{(\beta^{2}q)}+|q+\tanh(\beta^{2}q)|}.
\end{equation}
An alternative way to understand that the smaller eigenvalue could not be used to determine $\alpha_c$, is that
it leads to a non-physical solution $  \alpha_{c}=\frac{ \beta^{-4}  }{1+q\tanh{\beta^{2}q}-|q+\tanh{(\beta^{2}q)} | } $.
Because in a special case of large $\beta$ limit and positive $q$,
$\tanh{(\beta^{2}q)} \simeq     1-2e^{-2\beta^{2}q}  $,
then we have $\alpha_{c}\simeq\frac{ e^{2\beta^{2}q}   }{2(1-q)\beta^{4} }$, which implies that
this value tends to $\infty$ which is in contradiction with
the expectation that learning should be easier given noise-free data.

\section*{References}


\begin{thebibliography}{10}

\bibitem{Marr-1970}
D.~Marr.
\newblock {A Theory for Cerebral Neocortex}.
\newblock {\em Proceedings of the Royal Society of London B: Biological
  Sciences}, 176(1043):161--234, 1970.

\bibitem{Barlow-1989}
H.B. Barlow.
\newblock Unsupervised learning.
\newblock {\em Neural Computation}, 1:295--311, 1989.

\bibitem{Hinton-2006b}
G~Hinton, S~Osindero, and Y~Teh.
\newblock A fast learning algorithm for deep belief nets.
\newblock {\em Neural Computation}, 18:1527--1554, 2006.

\bibitem{Bengio-2008}
Nicolas Le~Roux and Yoshua Bengio.
\newblock Representational power of restricted boltzmann machines and deep
  belief networks.
\newblock {\em Neural Comput.}, 20(6):1631--1649, 2008.

\bibitem{Bara-2012}
Adriano Barra, Alberto Bernacchia, Enrica Santucci, and Pierluigi Contucci.
\newblock On the equivalence of hopfield networks and boltzmann machines.
\newblock {\em Neural Networks}, 34:1--9, 2012.

\bibitem{Huang-2015b}
Haiping Huang and Taro Toyoizumi.
\newblock Advanced mean-field theory of the restricted boltzmann machine.
\newblock {\em Phys. Rev. E}, 91:050101, 2015.

\bibitem{nips-2015}
Marylou Gabrie, Eric~W Tramel, and Florent Krzakala.
\newblock Training restricted boltzmann machine via the
  thouless-anderson-palmer free energy.
\newblock In C.~Cortes, N.~D. Lawrence, D.~D. Lee, M.~Sugiyama, and R.~Garnett,
  editors, {\em Advances in Neural Information Processing Systems 28}, pages
  640--648. Curran Associates, Inc., 2015.

\bibitem{Mezard-2017}
Marc M\'ezard.
\newblock Mean-field message-passing equations in the hopfield model and its
  generalizations.
\newblock {\em Phys. Rev. E}, 95:022117, 2017.

\bibitem{Monasson-2017}
J.~Tubiana and R.~Monasson.
\newblock Emergence of compositional representations in restricted boltzmann
  machines.
\newblock {\em Phys. Rev. Lett.}, 118:138301, 2017.

\bibitem{Song-2017}
J.~{Song}, M.~{Marsili}, and J.~{Jo}.
\newblock {Resolution and relevance trade-offs in deep learning}.
\newblock {\em J. Stat. Mech}, 2018:123406, 2018.

\bibitem{Decelle-2017}
A.~Decelle, G.~Fissore, and C.~Furtlehner.
\newblock {Spectral dynamics of learning in restricted Boltzmann machines}.
\newblock {\em EPL (Europhysics Letters)}, 119(6):60001, 2017.

\bibitem{Sala-2017}
Domingos S.~P. Salazar.
\newblock Nonequilibrium thermodynamics of restricted boltzmann machines.
\newblock {\em Phys. Rev. E}, 96:022131, 2017.

\bibitem{RSB-2018}
Gavin~S. Hartnett, Edward Parker, and Edward Geist.
\newblock Replica symmetry breaking in bipartite spin glasses and neural
  networks.
\newblock {\em Phys. Rev. E}, 98:022116, 2018.

\bibitem{icml-2008}
Tijmen Tieleman.
\newblock Training restricted boltzmann machines using approximations to the
  likelihood gradient.
\newblock In W.W. Cohen, A.~McCallum, and S.T. Roweis, editors, {\em
  Proceedings of the 25th International Conference on Machine Learning}, pages
  1064--1071. ACM, New York, NY, USA, 2008.

\bibitem{Huang-2016b}
Haiping Huang and Taro Toyoizumi.
\newblock Unsupervised feature learning from finite data by message passing:
  Discontinuous versus continuous phase transition.
\newblock {\em Phys. Rev. E}, 94:062310, 2016.

\bibitem{Huang-2017}
Haiping Huang.
\newblock {Statistical mechanics of unsupervised feature learning in a
  restricted Boltzmann machine with binary synapses}.
\newblock {\em Journal of Statistical Mechanics: Theory and Experiment},
  2017(5):053302, 2017.

\bibitem{Remi-2011}
S.~Cocco, R.~Monasson, and V.~Sessak.
\newblock High-dimensional inference with the generalized hopfield model:
  Principal component analysis and corrections.
\newblock {\em Phys. Rev. E}, 83:051123, 2011.

\bibitem{Barra-2017}
Adriano Barra, Giuseppe Genovese, Peter Sollich, and Daniele Tantari.
\newblock Phase transitions in restricted boltzmann machines with generic
  priors.
\newblock {\em Phys. Rev. E}, 96:042156, 2017.

\bibitem{Barra-2018}
Adriano Barra, Giuseppe Genovese, Peter Sollich, and Daniele Tantari.
\newblock Phase diagram of restricted boltzmann machines and generalized
  hopfield networks with arbitrary priors.
\newblock {\em Phys. Rev. E}, 97:022310, 2018.

\bibitem{Huang-2018}
Haiping Huang.
\newblock Role of zero synapses in unsupervised feature learning.
\newblock {\em Journal of Physics A: Mathematical and Theoretical}, 51:08LT01,
  2018.

\bibitem{DL-2016}
Ian Goodfellow, Yoshua Bengio, and Aaron Courville.
\newblock {\em Deep Learning}.
\newblock MIT Press, Cambridge, MA, 2016.

\bibitem{cavity-2001}
M.~M\'ezard and G.~Parisi.
\newblock The bethe lattice spin glass revisited.
\newblock {\em Eur. Phys. J. B}, 20:217, 2001.

\bibitem{MM-2009}
M.~M\'ezard and A.~Montanari.
\newblock {\em Information, Physics, and Computation}.
\newblock Oxford University Press, Oxford, 2009.

\bibitem{Yedidia-2005}
J.~S. Yedidia, W.~T. Freeman, and Y.~Weiss.
\newblock Constructing free energy approximations and generalized belief
  propagation algorithms.
\newblock {\em IEEE Trans Inf Theory}, 51:2282--2312, 2005.

\bibitem{Nishimori-2001}
H.~Nishimori.
\newblock {\em Statistical Physics of Spin Glasses and Information Processing:
  An Introduction}.
\newblock Oxford University Press, Oxford, 2001.

\bibitem{ICLR-17}
Pau {Rodr{\'\i}guez}, Jordi {Gonz{\`a}lez}, Guillem {Cucurull}, Josep~M.
  {Gonfaus}, and Xavier {Roca}.
\newblock {Regularizing CNNs with locally constrained decorrelations}.
\newblock {\em arXiv:1611.01967}, 2016.

\bibitem{dropcon}
Li~Wan, Matthew Zeiler, Sixin Zhang, Yann~L. Cun, and Rob Fergus.
\newblock Regularization of neural networks using dropconnect.
\newblock In Sanjoy Dasgupta and David Mcallester, editors, {\em Proceedings of
  the 30th International Conference on Machine Learning (ICML-13)}, volume~28,
  pages 1058--1066. JMLR Workshop and Conference Proceedings, 2013.

\bibitem{Huang-2018b}
Haiping Huang.
\newblock Mechanisms of dimensionality reduction and decorrelation in deep
  neural networks.
\newblock {\em Phys. Rev. E}, 98:062313, 2018.

\bibitem{Engel-1992}
A.~Engel, H.~M. K\"ohler, F.~Tschepke, H.~Vollmayr, and A.~Zippelius.
\newblock Storage capacity and learning algorithms for two-layer neural
  networks.
\newblock {\em Phys. Rev. A}, 45:7590--7609, 1992.

\bibitem{Barkai-1992}
E.~Barkai, D.~Hansel, and H.~Sompolinsky.
\newblock Broken symmetries in multilayered perceptrons.
\newblock {\em Phys. Rev. A}, 45:4146--4161, 1992.

\bibitem{Lake-2017}
Brenden~M. Lake, Tomer~D. Ullman, Joshua~B. Tenenbaum, and Samuel~J. Gershman.
\newblock Building machines that learn and think like people.
\newblock {\em Behavioral and Brain Sciences}, 40:e253, 2017.

\bibitem{Huang-2013}
Haiping Huang, K~Y~Michael Wong, and Yoshiyuki Kabashima.
\newblock Entropy landscape of solutions in the binary perceptron problem.
\newblock {\em Journal of Physics A: Mathematical and Theoretical}, 46:375002,
  2013.

\end{thebibliography}


\end{document}